\documentclass[amsmath,amssymb,11pt]{article}
\usepackage{jheppub2}
\pdfoutput=1
\usepackage{graphicx,epsfig}
\usepackage{float}
\usepackage{amsmath}
\usepackage{subfloat}
\usepackage[utf8]{inputenc}
\usepackage{hyperref}
\usepackage{caption}
\usepackage{mathrsfs}
\usepackage{subcaption} 
\usepackage{color}
\usepackage{overpic}
\usepackage{placeins}
\usepackage[dvipsnames]{xcolor}
\usepackage{physics}
\usepackage[multiple]{footmisc}
\usepackage{comment}
\usepackage{tensor}
\usepackage[normalem]{ulem}
\usepackage{orcidlink}
\usepackage{soul}
\usepackage{cancel}
\usepackage{graphicx}
\usepackage{subcaption}
\graphicspath{{Charged G H plots/}}
\newcommand{\cA}{\mathcal A}

\newcommand{\eps}{\epsilon}
\newcommand{\ren}{\mathrm{ren}}

\newcommand{\grav}{\mathrm{grav}}

\usepackage{pifont}

\newcommand{\beq}{\begin{equation}}
\newcommand{\eeq}{\end{equation}}

\newcommand{\be}{\begin{equation}}
\newcommand{\ee}{\end{equation}}
\newcommand{\ba}{\begin{eqnarray}}
\newcommand{\ea}{\end{eqnarray}}

\newcommand{\cF}{\mathcal{F}}
\newcommand{\cFs}{\mathcal{F}_{\rm s}}

\newcommand{\beqa}{\begin{eqnarray}}
\newcommand{\eeqa}{\end{eqnarray}}

\usepackage{tikz}
\usetikzlibrary{positioning}
\usetikzlibrary{intersections}
\usetikzlibrary{fadings} 
\usetikzlibrary{arrows.meta} 
\usetikzlibrary{arrows}

\tikzfading[name=fade out,
inner color=transparent!0,
outer color=transparent!100]

\definecolor{cherryblossompink}{rgb}{1.0, 0.72, 0.77}
\definecolor{lightblue}{rgb}{0.68, 0.85, 0.9}

\usetikzlibrary{decorations.pathmorphing}
\usetikzlibrary{decorations.pathreplacing,decorations.markings}

\usetikzlibrary{backgrounds,automata}

\begin{document}

\title{Charged quantum solitons}

\author{Yuvraj Dhunna,}
\author{Robie A.~Hennigar,}
\author{Ayan K. Patra,}
\author{and Simon F. Ross}
\affiliation{Centre for Particle Theory, Department of Mathematical Sciences, Durham University, Durham DH1 3LE, UK}
\emailAdd{yuvraj.dhunna@durham.ac.uk, robie.a.hennigar@durham.ac.uk, ayan-kumar.patra@durham.ac.uk, s.f.ross@durham.ac.uk}

\abstract{We extend our recent work on the construction of horizonless solutions describing the quantum backreaction of thermal fields in AdS$_3$ to include charged matter. The solutions are obtained from branes in charged C-metric solutions. We show that the charged C-metric goes into itself under a double analytic continuation, under which our solutions are related to the previously studied charged quantum BTZ solutions. Unlike in the uncharged case, for solutions which have a horizon before backreaction, we have backreacted solutions where the temperature of the thermal matter is either above or below the temperature of the horizon.}

\maketitle

\section{Introduction}

Holography provides a dual gravitational description of strongly-coupled quantum field theories. Introducing braneworld boundaries~\cite{Karch:2000ct} enables us to dynamically couple the strongly-coupled field theory to gravity, providing an approach to calculating the semi-classical backreaction of the matter fields~\cite{Emparan:2002px}. This provides a unique non-perturbative method for understanding such quantum backreaction. This programme has been carried out in practice for branes in C-metric geometries, starting with \cite{Emparan:1999wa,Emparan:1999fd}, see \cite{Panella:2024sor} for a review. The C-metric is a four-dimensional solution, with a black hole in the bulk. A large part of the past work has considered branes which intersect the bulk black hole horizon, so the induced geometry on the brane is also a black hole, describing a backreacted black hole solution in the three-dimensional theory, which is referred to as the quantum BTZ black hole \cite{Emparan:1999fd,Emparan:2002px,Emparan:2020znc}. These constructions now include rotating quantum black holes~\cite{Emparan:2020znc}, charged and rotating quantum black holes~\cite{Feng:2024uia,Climent:2024nuj,Bhattacharya:2025tdn, Mensikova:2026ejy}, and quantum black holes with de Sitter asymptotics~\cite{Emparan:2022ijy, Panella:2023lsi,Climent:2024wol}.

The availability of exact solutions makes this framework particularly valuable for exploring questions in semiclassical gravity beyond the regime of perturbatively small backreaction. These solutions have been used to investigate cosmic censorship~\cite{Emparan:2020rnp,Frassino:2025buh}, black hole singularities~\cite{Kolanowski:2023hvh}, and quantum versions of the Penrose and reverse isoperimetric inequalities~\cite{Frassino:2024bjg}. More generally, braneworld holography has supplied a higher-dimensional origin for extended black hole thermodynamics~\cite{Frassino:2022zaz} and enabled studies of thermodynamic stability, phase transitions and critical behaviour~\cite{Johnson:2023dtf,Frassino:2023wpc,HosseiniMansoori:2024bfi,Cao:2026jls}.  Analyses of quasinormal modes have revealed how quantum effects modify relaxation and dynamical stability~\cite{Cartwright:2024iwc,Cartwright:2025fay}, including quantum-induced superradiance, and shown how the structure of black hole singularities can be extracted from the asymptotic quasinormal spectrum~\cite{Coviello:2026tol,Coviello:2026hdc}.

In recent work \cite{Hennigar:2026mxv}, we considered instead the theory on a brane which does not intersect the horizon. The bulk black hole is then dual to a thermal gas on the brane, and the induced metric on the brane is a horizonless solution describing the quantum backreaction of this thermal gas, which we dubbed the quantum soliton. This name reflects the fact that this quantum soliton geometry is related to the BTZ black hole by a double analytic continuation, similar to the relation between the classical AdS soliton \cite{Horowitz:1998ha} and the planar AdS black hole. Our holographic quantum solitons share some features with analogous solutions obtained in perturbative semi-classical gravity~\cite{Thompson:2025kfm, Thompson:2025jkn}, though those have different boundary conditions on the quantum fields. In the present paper, we extend this discussion of the quantum solitons to charged C-metric solutions in the bulk, dual to considering the quantum backreaction of charged matter. We find that the qualitative features of our previous study carry through in this case, with some interesting new features. 

In particular, we show that the charged C-metric goes into itself under a double analytic continuation $t \to i \varphi$, $\phi \to i \tau$. This exchanges the quantum BTZ brane and the soliton brane. The C-metric has a zero acceleration limit where it reduces to the charged Reissner-Nordstr\"om AdS black hole; this is the zero backreaction limit for the quantum soliton. There are charged black holes with spherical, planar and hyperbolic horizons. We show that in the hyperbolic case, there are regular C-metric solutions where the mass parameter $\mu$ is negative, unlike in the uncharged case, where the limiting hyperbolic black hole was regular for $\mu <0$ but the C-metric was not. Thus, we can use the quantum soliton brane to describe the backreaction in the hyperbolic case for both signs of the mass parameter. The solutions with negative mass have a temperature of the thermal matter below the temperature of the horizon; this is also true for some solutions with positive mass. Thus, in the charged case we can study backreaction in cases where the temperature of the thermal matter is either above or below the temperature of the horizon. 

For the quantum soliton brane, we find that in the regime of small backreaction, where the brane effective theory is valid, the backreaction is dominated by the effect of the mass; charge introduces a subleading correction. Thus, the quantitative features of the backreaction in this charged case reduce to the uncharged case analysed in our previous paper \cite{Hennigar:2026mxv}. This is because of a sub-extremality constraint on the charge and mass parameters required for the existence of a bulk black hole horizon. In the small backreaction regime for the quantum soliton, this is a non-linear constraint. This is quite different from the quantum BTZ case considered in \cite{Feng:2024uia,Climent:2024nuj}, where the extremality constraint is linear in the small backreaction regime. 

In the next section, we begin by considering the Reissner-Nordstrom AdS$_4$ black hole and its double analytic continuation, a bulk soliton. These are the solutions obtained from the C-metric in the zero backreaction limits for the two kinds of branes. We explain how the boundaries of these solutions are written in a locally AdS$_3$ frame: these will be the zero backreaction limits of the branes. In section \ref{sec:Cmetric}, we turn to the charged C-metric, and describe the different possible pure tension branes that can be introduced in the bulk. In section \ref{sec:qsol}, we summarize the physics of the backreacted quantum soliton solutions, describing the backreaction of thermal matter on the geometry. In section \ref{sec:thermo}, we briefly discuss the thermodynamics, showing that the quantum soliton satisfies a first law as in the uncharged case. We briefly discuss the supersymmetry of these solutions in appendix \ref{sec:susy}. The supersymmetry condition for the C-metric in our coordinate system is simplified compared to previous analyses \cite{Klemm:2013eca,Ferrero:2020twa}. We note that in this charged but non-rotating case the only supersymmetric C-metric solutions are naked singularities. A regular solution found in \cite{Ferrero:2020twa} is seen to correspond to the limiting case where the C-metric reduces to the bulk soliton. None of the brane configurations preserve supersymmetry. Details of the action calculation for the thermodynamics are given in appendix \ref{action}.

\section{Unbackreacted solutions}

In this section we describe the four-dimensional solutions whose conformal boundaries are the zero-backreaction limits of the brane geometries studied in the rest of the paper, together with the CFT states they describe. There are four cases in total, corresponding to global AdS$_3$, AdS-Rindler, BTZ black hole, and conical singularity backgrounds. These are obtained from the boundary geometries of Reissner--Nordstr{\"o}m--AdS black holes and their double analytical continuations. 

\subsection{Bulk theory and holographic dictionary}

We study four-dimensional Einstein--Maxwell theory with negative cosmological constant, 
\begin{equation}
I
=
\frac{1}{16\pi G_4}
\int \mathrm{d}^4x\,\sqrt{-g}
\left(
R+\frac{6}{\ell_4^2}
-\frac{\ell_\star^2}{4}F_{\mu\nu}F^{\mu\nu}
\right),
\qquad
\ell_\star^2=\frac{16\pi G_4}{g_\star^2} \,, 
\end{equation}
where $F = {\rm d} \mathcal{A}$ and $g_\star$ is the dimensionless gauge coupling. The equations of motion are
\begin{align}
G_{\mu\nu}
-\frac{3}{\ell_4^2}g_{\mu\nu}
&=
\frac{\ell_\star^2}{2}
\left(
F_{\mu\rho}F_{\nu}{}^{\rho}
-\frac{1}{4}g_{\mu\nu}
F_{\rho\sigma}F^{\rho\sigma}
\right),
\label{einstein_equation}
\\
\nabla_\mu F^{\mu\nu}&=0.
\label{maxwell_equation}
\end{align}
The parameters in the bulk action are related to the central charges characterizing the CFT$_3$, which appear in the two-point functions of the stress tensor and the current operator,
\be 
c_3 = \frac{\ell_4^2}{G_4} \,, \qquad c_J = \frac{1}{g_\star^2}, \quad \text{so that} \quad \frac{\ell_4}{\ell_\star} = \sqrt{\frac{c_3}{16 \pi c_J}} \, .
\ee

In what follows we will repeatedly make use of some standard holographic technology~\cite{deHaro:2000vlm,Skenderis:2002wp}, and we present this here to fix our conventions. Near the conformal boundary we use Fefferman--Graham coordinates, 
\begin{align}
    {\rm d}s^2 &= \frac{\ell_4^2}{z^2} \left( {\rm d}z^2 + g_{ij}(z, x) {\rm d}x^i {\rm d}x^j\right) \, ,\quad g_{ij} = g_{ij}^{(0)} + z^2 g_{ij}^{(2)} + z^3 g_{ij}^{(3)} + \cdots \,, 
    \\
    \mathcal{A}_i &= \mathcal{A}^{(0)}_i + z \mathcal{A}^{(1)}_i + \cdots \, ,
\end{align}
and we work in a radial gauge where $\mathcal{A}_z = 0$. The boundary metric $g^{(0)}$ and the gauge field $\mathcal{A}^{(0)}$ are sources, while the one-point functions of the stress tensor and current are
\be 
\left<T_{ij}\right> = \frac{3 c_3}{16 \pi} g_{ij}^{(3)} \,, \qquad \left<J_i\right> = c_J \mathcal{A}_i^{(1)} \, .
\ee
The current can be equivalently written in terms of the component of the electric flux that is normal to the boundary. If we let $n^\mu$ be the normal pointing toward the boundary and take $\Omega$ to be defined by writing the metric on a cutoff surface as $\Omega^2 g^{(0)}$, then the current is
\be 
\left<J^i\right>
 = - c_J \lim_{\Omega\to\infty} \Omega^3 n_\mu F^{\mu i} \, .
\ee
This is the most convenient form for computing the current. 

The boundary gauge field $\mathcal{A}^{(0)}$ is a non-dynamical background field for the $U(1)$ symmetry of the dual CFT. From it we can extract gauge-invariant content such as its field strength $F^{(0)} = {\rm d} \mathcal{A}^{(0)}$ and its holonomies around closed cycles. The current $\left<J^i\right>$ is the CFT response, and it is related to the conserved charge via
\be 
Q = - \int_{\Sigma} {\rm d}^2 x \sqrt{\sigma } u_i \left<J^i\right> \, ,
\ee
where $\Sigma$ is a Cauchy slice of the boundary with future-pointing unit timelike normal $u^i$. We shall also study integrals of $F$ and $\star F$ over surfaces in the bulk that do not reach the boundary --- these will be referred to as fluxes. 

Our analysis will repeatedly make use of Weyl transformations of the boundary metric, $g^{(0)} \to \Omega^2 g^{(0)}$. There is no Weyl anomaly in three dimensions, so the stress tensor and current transform as
\be 
\left<T_i{}^j\right> \to \Omega^{-3} \left<T_i{}^j\right> \,, \qquad \left<J^i\right> \to \Omega^{-3} \left<J^i\right>  \,,
\ee
while the source $\mathcal{A}^{(0)}$ is Weyl invariant. In particular the densities $\sqrt{-g^{(0)}} \left<T_i{}^j\right>$ and $\sqrt{-g^{(0)}} \left<J^i\right>$ are invariant, and therefore the conserved charges are invariant as well. 

\subsection{Charged black holes}

The solutions of interest are the Reissner--Nordstr{\"o}m--AdS$_4$ solution and its double analytic continuation. For Reissner--Nordstr{\"o}m--AdS$_4$ the metric is~\cite{Brill:1997mf}
\begin{align} \label{rnAdS}
    \dd s^2 &= - \ell_4^2 \cF(\rho) \dd \tau^2 + \frac{\dd \rho^2}{\cF(\rho)} + \rho^2 \dd \Omega_\kappa^2 \, ,
    \qquad 
    \cF(\rho) = \kappa + \frac{\rho^2}{\ell_4^2} - \frac{\hat{\mu} \ell_4}{\rho} + \frac{\hat{q}^2 \ell_4^2}{\rho^2} \, ,
    \\
    \dd \Omega^2_\kappa &= \frac{\dd x^2}{1-\kappa x^2} + (1-\kappa x^2) \dd \phi^2 \, .
    \label{eq:omega_kappa}
\end{align}
Here we have taken a factor of $\ell_4$ from the standard Schwarzschild time coordinate to make it dimensionless, the parameters $\hat{\mu}$, $\hat{q}$ are dimensionless mass and charge parameters, and we can choose coordinates so that $\kappa = \pm 1,0$. 
For $\kappa = +1$, we can rewrite $\dd \Omega_\kappa$ as the metric on the round unit sphere by setting $x = \cos \theta$. Thus the range of $x$ is $x \in [-1,1]$. For $\kappa = -1$, we can rewrite $\dd \Omega_\kappa$ as a metric on $\mathbb{H}^2$ by setting $x = \sinh  \theta$. This form of the metric covers half the hyperbolic disc (see \cite{Hennigar:2026mxv} for details); it will be convenient for our later purposes. The planar case $\kappa = 0$ can be obtained as a limit and will not be discussed separately. 

The gauge field of the solution reads 
\be 
\mathcal{A} = \frac{2 \ell_4}{\ell_\star} \left[\hat{e} \ell_4 \left(\frac{1}{\rho_0} - \frac{1}{\rho} \right) \dd \tau + \hat{g} \left(x-x_\kappa \right) \dd \phi \right] \,,\qquad \hat{q}^2 = \hat{e}^2 + \hat{g}^2 \, ,
\ee
so $\hat{e}$ is the electric and $\hat{g}$ the magnetic charge parameter. The additive constants in the gauge potential are chosen for regularity. The $\dd \tau $ component vanishes at the horizon $\rho = \rho_0$,  the largest root of $\cF$, as required for regularity of the Euclidean section. For $\kappa = +1$ we take $x_{+1} = 1$, so that the potential is regular in a patch containing the north pole $x = 1$. For $\kappa = -1$ there is no axis in the patch covered by our coordinates and $x_{-1} = 0$ is a convenient choice. 

Writing $u_0 = \rho_0/\ell_4$, the temperature conjugate to $\tau$ and the potential difference between infinity and the horizon are
\be \label{eq:temp-chemical-potential}
T_{\rm H} = \frac{\ell_4 \cF'(\rho_0)}{4 \pi } = \frac{3 u_0^4 + \kappa u_0^2 - \hat{q}^2}{4 \pi u_0^3} \,, \qquad \mu_e = \mathcal{A}_\tau \big|_{\rho \to \infty} = \frac{2 \ell_4 \hat{e}}{\ell_\star u_0} \, .
\ee

\begin{figure}
    \centering
    \includegraphics[width=\linewidth]{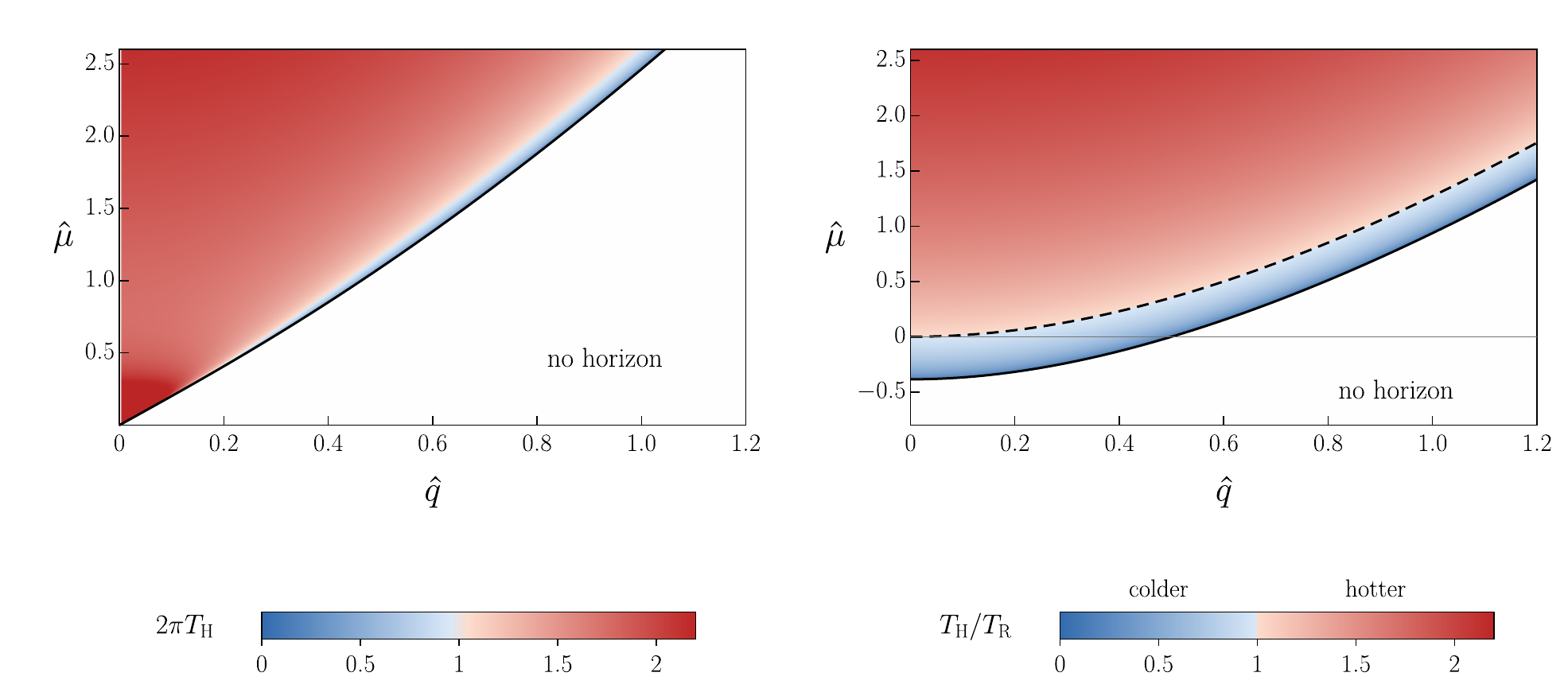}
    \caption{Parameters $(\hat{q}, \hat{\mu})$ for which the RN-AdS black hole has a horizon (shaded regions), bounded by the extremal solutions (solid, black curves). The left plot shows $\kappa = +1$, and the right plot $\kappa = -1$. For the case $\kappa = -1$, the parameter space is further split into two regions, demarcated by the dashed, black curve corresponding to when the black hole horizon temperature $T_{\rm H}$ is the same  as the boundary Rindler horizon $T_{\rm R} = 1/(2 \pi)$. In both plots, shading indicates temperature, with warmer colours corresponding to higher temperature. }
    \label{fig:rn-params}
\end{figure}

\paragraph{Existence of a bulk horizon.} 
The function $\cF$ has a positive root if and only if $\hat{\mu} \ge \hat{\mu}_{\rm ext}(\hat{q})$, where $\hat{\mu}_{\rm ext}(\hat{q})$ is the extremal value. The extremal value is determined by solving $\cF=0$ and $\cF'=0$. These are easily solved to give $\hat \mu$, $\hat q$ as functions of the extremal horizon radius $u_\star$, 
\be 
\hat{\mu}_{\rm ext} = 2 u_\star \left(2 u_\star^2 + \kappa \right) \,, \quad \hat q^2 = u_\star^2 (3 u_\star^2 + \kappa)  \, ,
\ee
which can be solved to explicitly determine
\be \label{kpbound}
\hat{\mu}^2_{\rm ext} = \frac{2}{27} \left[\kappa (36 \hat{q}^2 - 1) + (1+12 \hat{q}^2)^{3/2} \right] \, .
\ee
We show the parameter space of the allowed solutions in Figure~\ref{fig:rn-params}. There are two important features of the extremal horizon relationship:
\begin{itemize}
    \item For $\kappa = -1$, the extremal mass parameter is negative for $\hat{q} < 1/2$. There are therefore black holes with negative mass parameter in the range $\hat{\mu}_{\rm ext} \le \hat{\mu} \le 0$. This window shrinks to zero as $\hat{q} \to 1/2$. 
    \item For either sign of $\kappa$, at large $\hat{q}$ the bound becomes
    \be \label{largeqconst}
    \hat{\mu} \gtrsim \frac{4}{3^{3/4}} \hat{q}^{3/2} \, .
    \ee
\end{itemize}

\subsection{Double analytic continuation: charged AdS solitons}

The black hole~\eqref{rnAdS} has a double analytic continuation to a bulk soliton.\footnote{The $\kappa=0$ analogue of this soliton has been extensively discussed, see e.g. \cite{Anabalon:2021tua}.} In the gauge field the continuation also exchanges the roles of the electric and magnetic parameters:
\be 
\tau \to i \Phi \,, \quad \phi \to i T \,, \quad x \to \bar{y} \,, \quad \hat{\mu} \to \mu \,, \quad \hat{e} \to i g \,, \quad \hat{g} \to -i e \, .
\ee
The factors of $i$ ensure a real gauge field and imply that $\hat{q}^2 \to - q^2$, with $q^2 = e^2 + g^2$. Up to a constant shift of the $T$ component of the potential, the result is
\begin{align} \label{solads}
    \dd s^2 &= \ell_4^2 \cFs(\rho) \dd \Phi^2 + \frac{\dd \rho^2}{\cFs(\rho)} + \rho^2 \left[\frac{\dd \bar{y}^2}{1 - \kappa \bar{y}^2} - \left(1- \kappa \bar{y}^2 \right) \dd T^2\right] \,, 
    \\
    \cFs(\rho) &= \kappa + \frac{\rho^2}{\ell_4^2} - \frac{\mu \ell_4}{\rho} - \frac{q^2 \ell_4^2}{\rho^2 } \,, \qquad \mathcal{A} = \frac{2 \ell_4}{\ell_\star} \left[g \ell_4 \left(\frac{1}{\rho} - \frac{1}{\rho_0} \right) \dd \Phi + e (\bar{y}- \bar {y}_\kappa)\, \dd T\right] \, .
\end{align}
For $\kappa = +1$ the two-dimensional $(\bar{y}, T)$ sector is the static patch of dS$_2$, with $-1 < \bar{y} < 1$ and Killing horizons at $\bar{y} = \pm 1$. For $\kappa = -1$, it is global AdS$_2$, with $\bar{y} \in \mathbb{R}$. The radial direction terminates at the largest positive root $\rho_0$ of $\cFs$, where the $\Phi$ circle pinches off. We restrict here to simple roots with $\cFs'(\rho_0) > 0$ and impose
\be 
\Phi \sim \Phi + \Delta \Phi \,, \qquad \Delta \Phi = \frac{4 \pi }{\ell_4 \cFs'(\rho_0)} \, ,
\ee
to ensure regularity. For $q \neq 0$ a positive root always exists because $\cFs \to -\infty$ as $\rho \to 0$ and is positive at large $\rho$. Thus, for the soliton there is no analogue of the sub-extremality condition in the black hole. However, a double root of $\cF$ is still possible for $\kappa=-1$. This degenerate case needs a separate treatment, as it has an infinite throat; we will not consider it further. The gauge field potential has been chosen such that the $\dd \Phi$ component is regular where the circle pinches off. We choose $\bar y_{+1} = 1$ to make the gauge field regular at the horizon at $\bar{y} = 1$ and $\bar y_{-1} =0$, as in the black hole case.

\subsection{Thermal CFTs on AdS$_3$ and AdS-Rindler}
\label{sec:ads-rindler}

Let us now examine the boundary states dual to these bulk solutions, showing how the boundary geometry is related to the forms that appear in the zero-backreaction limit of the C-metric solutions presented later. We consider first the boundary of the black hole \eqref{rnAdS}, which will give the unbackreacted version of the quantum solitons. 

We take as a representative of the boundary geometry
\be \label{eq:bdry_mtrc_rnads}
\dd s^2_\partial = \ell_3^2 \left(-\dd \tau^2 + \dd \Omega_\kappa^2 \right) \, .
\ee
For $\kappa = +1$ this is the Einstein static universe $\mathbb{R} \times S^2$ of radius $\ell_3$. For $\kappa = -1$ the boundary is $\mathbb{R} \times \mathbb{H}^2$. The scale $\ell_3$ is arbitrary, since the bulk fixes only the conformal class. The holographic boundary data are
\begin{align}
    \mathcal{A}^{(0)} = \mu_e \dd \tau + \frac{2 \hat{g} \ell_4}{\ell_\star} (x- x_\kappa) \dd \phi \,, \qquad F^{(0)} = \frac{2 \hat{g} \ell_4}{\ell_\star} \dd x \wedge \dd \phi \,,
    \\
    \left<T_i{}^j\right> = \frac{\hat{\mu} c_3}{16 \pi \ell_3^3} \, {\rm diag} (-2,1,1) \, , \qquad \left<J^i\right> = \frac{2 c_J \hat{e} \ell_4}{\ell_\star \ell_3^3} \delta_\tau^i \, .
\end{align}
The CFT is in a thermal state at temperature $T_{\rm H}$ with chemical potential $\mu_e$ in a static background magnetic field. The two bulk charge parameters play different roles in the boundary interpretation. The parameter $\hat{e}$ is a state parameter, setting the chemical potential $\mu_e$, the holonomy of $\mathcal{A}^{(0)}$, and the charge density. The parameter $\hat{g}$ is purely a source, providing the background field $F^{(0)}$~\cite{Hartnoll:2007ai}. It has no corresponding response. 

For $\kappa = +1$ the charges and energy are finite and read
\be \label{eq:rnads-charge-energy}
Q_e = 2 \sqrt{\pi c_3 c_J} \, \hat{e} \, , \qquad Q_m = c_J \int_{S^2} F^{(0)} = 2 \sqrt{\pi c_3 c_J} \, \hat{g} \,, \quad E_\tau =\frac{\hat{\mu} c_3}{2} \, ,
\ee
where $E_\tau$ is the energy associated to the timelike Killing vector $\partial_\tau$. Here $Q_m$ represents a flux of the background field, rather than a charge of the CFT state. In the case $\kappa = -1$, the area is infinite and one should instead think of the quantities as densities. 

\paragraph{Weyl transform to AdS$_3$ and Rinder-AdS$_3$.} To make contact with the quantum solitons to be discussed below, we consider a region of the boundary $x > 0$ and do a coordinate and Weyl transformation
\be 
x = \frac{\ell_3}{r} \,, \qquad \dd s_3^2 = \Omega^2 \dd s_\partial^2 \,, \qquad \Omega = \frac{1}{x} = \frac{r}{\ell_3}\, .
\ee
This transformation maps the region $x > 0$ of~\eqref{eq:bdry_mtrc_rnads} to
\be \label{eq:zero-backreaction-qsol}
\dd s_3^2 = - r^2 \dd \tau^2 + \frac{\dd r^2}{f(r)} + \ell_3^2 f(r) \dd \phi^2 \, , \quad f(r) = \frac{r^2}{\ell_3^2} - \kappa \, .
\ee
For $\kappa = +1$ this is global AdS$_3$ with $r \in [\ell_3, \infty)$. The $\phi$ circle closes off smoothly at $r = \ell_3$ with period $\phi \sim \phi + 2 \pi $.\footnote{Setting $r^2 = \ell_3^2 + R^2$ and $t = \ell_3 \tau$ this metric becomes $\dd s_3^2 = - (1+R^2/\ell_3^2) \dd t^2 + \dd R^2/(1+R^2/\ell_3^2) + R^2 \dd \phi^2$. } For $\kappa = -1$ this is Rindler-AdS$_3$ with $r \in (0, \infty)$ and $r = 0$ (which corresponds to $x \to \infty$, being the asymptotic region of $\mathbb{H}^2$) is a Killing horizon of $\partial_\tau$. The horizon has Rindler temperature $T_{\rm R} = 1/(2 \pi)$ with respect to $\partial_\tau$. 

In this frame the holographic data become
\begin{align}
\langle T_i{}^j\rangle_{\rm AdS}&=\frac{\hat\mu\,c_3}{16\pi r^3}\,{\rm diag}(-2,1,1),
\qquad
\langle J^i\rangle_{\rm AdS}=\frac{2c_J\,\hat e\,\ell_4}{\ell_\star\,r^3}\,\delta^i_\tau,
\label{adsbdyTJ}\\
\mathcal A^{(0)}&=\mu_e\,\dd\tau+\frac{2\hat g\ell_4}{\ell_\star}\left(\frac{\ell_3}{r}-x_\kappa\right)\dd\phi,
\qquad
F^{(0)}=-\frac{2\hat g\,\ell_4\ell_3}{\ell_\star\,r^2}\,\dd r\wedge\dd\phi \, .
\label{eq:ads_A0}
\end{align}

\begin{figure}
    \centering
    \includegraphics[width=\linewidth]{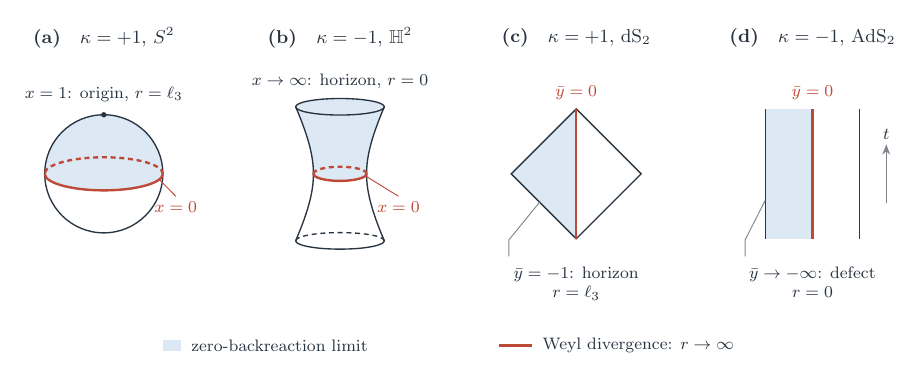}
    \caption{The two-dimensional factor of the boundary for the four different cases. The left two pictures (a), (b) are the boundaries of the black holes, and the right two pictures (c), (d) are the boundaries of the bulk soliton. The latter diagrams are Penrose diagrams, with time running upward. The shaded regions are those approached by the brane in the zero-backreaction limit of the C-metric solutions. From the boundary point of view, these are related to the standard boundary coordinates by a Weyl transformation, as described in the main text.
    The Weyl factor diverges at the red curves ($x = 0$ or $\bar{y} = 0$), which becomes the conformal boundary $r \to \infty$ of AdS$_3$ in the metric on the brane. 
    }
    \label{fig:two-copies-plots}
\end{figure}

\paragraph{Two boundary copies.} The Weyl factor $\Omega = 1/|x|$ diverges at $x = 0$, which becomes the conformal boundary at $r \to \infty$ in~\eqref{eq:zero-backreaction-qsol}. In obtaining~\eqref{eq:zero-backreaction-qsol} we restricted to $x > 0$. The region $x < 0$, with $x = -\ell_3/r$, gives a second copy of global AdS$_3$ for $\kappa = +1$, or a second AdS-Rindler exterior for $\kappa = -1$. The CFT on the original boundary thus gives a state on these two regions coupled through their common conformal boundary by transparent boundary conditions --- see~\cite{Hubeny:2009rc} for related considerations. We illustrate the two regions in the first two panels of Figure~\ref{fig:two-copies-plots}. The CFT state has $\left<T^r{}_\tau \right> = \left< J^r\right> = 0$, so there is no flux of energy or charge between the two copies. For $\kappa = +1$, each region carries half the total charge and energy in~\eqref{eq:rnads-charge-energy}.

\paragraph{Regularity and the horizon temperature.} For $\kappa = +1$, the stress tensor, current and gauge field are smooth at the origin $r = \ell_3$. For $\kappa = -1$, the Weyl factor vanishes at the Rindler-AdS horizon $r = 0$, compressing the asymptotic region of $\mathbb{H}^2$ onto it. The stress tensor components behave as $\hat{\mu}/r^3$, while the norm of the current and the boundary field strength behave as $\hat{e}^2/r^4$ and $\hat{g}^2/r^4$. The CFT data restricted to this chart is therefore fully regular only for $\hat{\mu} = \hat{e} = \hat{g} = 0$, which corresponds to AdS$_4$ written in a hyperbolic slicing. For all other parameters, at least one of the stress tensor, current, or background field is singular on the horizon. 

In the $\kappa = -1$ case, the physical interpretation is that of a CFT state of temperature $T_{\rm H}$ on an AdS-Rindler background of temperature $T_{\rm R}$ where the two temperatures generically \textit{disagree}. In the uncharged case, one has generically $T_{\rm H} \ge T_{\rm R}$ for $\hat{\mu} > 0$, with equality only when the bulk is vacuum AdS$_4$. The situation is more interesting with charge, and it is possible to have $T_{\rm H} < T_{\rm R}$ for positive mass as well. More explicitly, setting $T_{\rm H} = T_{\rm R}$ in~\eqref{eq:temp-chemical-potential} gives
\be 
\hat{q}^2 = u_0^2(u_0-1)(3u_0+1) \, , \qquad \hat{\mu}_{\rm R} = 2 u_0 (u_0-1)(2u_0 + 1) \,, \qquad u_0 \ge 1 \, . 
\ee
This is the dashed curve illustrated in Figure~\ref{fig:rn-params}. Every negative mass CFT state is colder than the horizon, and for $\hat{q} \neq 0$ the colder states also include a band with $0 \le \hat{\mu} \le \hat{\mu}_{\rm R}$.

\subsection{CFTs on BTZ and conical AdS$_3$}

We now turn to a study of the boundary geometry of the charged AdS solitons. A similar study for neutral bulk solutions was first carried out in~\cite{Hubeny:2009rc}, with a remarks on the charged generalization appearing in~\cite{Climent:2024nuj}.

\paragraph{Boundary geometry and CFT data.}
We take here as our representative of the conformal boundary
\be 
\dd s^2_\partial = \ell_3^2 \left[\dd \Phi^2 + \frac{\dd \bar{y}^2}{1 - \kappa \bar{y}^2} - \left(1-\kappa \bar{y}^2\right) \dd T^2 \right] \, .
\ee
This is dS$_2 \times S^1$ for $\kappa = +1$ and AdS$_2 \times S^1$ for $\kappa = -1$. The boundary data are
\begin{align}
\mathcal A^{(0)}&=-\frac{2g\ell_4^2}{\ell_\star\rho_0}\dd\Phi+
\frac{2e\ell_4}{\ell_\star}\bar y\dd T,
&F^{(0)}&=\frac{2e\ell_4}{\ell_\star}\dd\bar y\wedge\dd T,
\label{eq:soliton_sources}\\
\langle T_i{}^j\rangle&=\frac{\mu c_3}{16\pi\ell_3^3}\operatorname{diag}(1,1,-2),
&\langle J^i\rangle&=\frac{2c_Jg\ell_4}{\ell_\star\ell_3^3}\delta^i_\Phi,
\label{eq:soliton_vevs}
\end{align}
with the components ordered as $(T, \bar{y}, \Phi)$. The parameter $g$ determines a spatial holonomy 
\be 
W = \oint \mathcal{A}^{(0)} = - \frac{2 g \ell_4^2}{\ell_\star \rho_0} \Delta \Phi ,
\ee
and a current around that circle. Enforcing regularity in the bulk at $\rho = \rho_0$ then relates $W$ to the response. The parameter $e$ determines a background electric field, while the charge density $\left<J^T\right>$ vanishes. The state therefore carries no $U(1)$ charge, despite nonzero gauge field. For $\mu > 0$ the energy density of the state $- \left<T_T{}^T\right>$ is negative, but this is familiar behaviour of AdS solitons. Despite vanishing CFT charge, there are nontrivial bulk fluxes through the disc $D$ spanned by $(\rho, \Phi)$ at fixed $(\bar{y}, T)$,
\be 
\int_D F = - \frac{2 g \ell_4^2}{\ell_\star \rho_0} \Delta \Phi = W \,, \qquad c_J \int_D \star F = \frac{2 c_J e \ell_4^2}{\ell_\star \rho_0} \Delta \Phi \, .
\ee
These should not be interpreted as properties of the CFT state, but something external to it. In~\cite{Climent:2024nuj} these fluxes were interpreted as charges associated to a defect. 

\paragraph{Weyl transform to BTZ and conical AdS$_3$.} For $\bar{y} < 0$ let us set
\be 
\bar{y} = - \frac{\ell_3}{r} \, , \qquad \Omega = \frac{1}{|\bar{y}|} = \frac{r}{\ell_3} \,, \qquad \tilde{t} = \ell_3 T \, .
\ee
The Weyl rescaled metric $\dd s_3^2 = \Omega^2 \dd s^2_\partial$ is 
\be \label{solbdy}
\dd s_3^2 = - f(r) \dd \tilde{t}^2 + \frac{\dd r^2}{f(r)} + r^2 \dd \Phi^2 \,, \quad f(r) = \frac{r^2}{\ell_3^2} - \kappa \, .
\ee
For $\kappa = +1$, this is the BTZ exterior $r > \ell_3$, with the de Sitter horizon $\bar{y} = -1$ mapped to $r = \ell_3$. Rescaling to a $2\pi$-periodic angle, together with the corresponding radial and time rescalings, gives the usual BTZ horizon radius $r_+ = \ell_3 \Delta \Phi/(2 \pi)$. For $\kappa = -1$, the AdS$_2$ boundary $\bar{y} \to - \infty$ maps to $r = 0$. The metric has there a conical deficit $2 \pi - \Delta \Phi$, or a conical excess if $\Delta \Phi > 2 \pi$. 

In this frame the CFT data are
\begin{align}
    \left<T_i{}^j\right>_{\rm BTZ} &= \frac{\mu c_3}{16 \pi r^3} \, {\rm diag} (1,1,-2) \,, \quad &\left<J^i\right>_{\rm BTZ}& = \frac{2 c_J g \ell_4}{\ell_\star r^3} \delta_\Phi^i \, ,
    \\
    \mathcal{A}^{(0)} &= \frac{W}{\Delta \Phi} \dd \Phi + \frac{2 e \ell_4}{\ell_\star} \left(\frac{\delta_{\kappa, 1}}{\ell_3} - \frac{1}{r}\right) \dd \tilde{t} \, , \quad &F^{(0)}& = \frac{2 e \ell_4}{\ell_\star r^2} \dd r \wedge \dd \tilde{t} \, ,
\end{align}
with the components ordered $(\tilde{t}, r, \Phi)$. For $\kappa = +1$, the time component of the potential vanishes at the BTZ horizon and the electrostatic potential difference is $\mu_{\rm BTZ} = 2 e \ell_4/(\ell_\star \ell_3)$. Combined with the fact that $\left<T_t{}^t\right>=\left<T_r{}^r\right>$, the state is therefore regular across the BTZ horizon and coincides with the Hartle--Hawking-like state inherited from the smooth continuation through the de Sitter horizon. For $\kappa = -1$, nonzero parameters generically give rise to singular data at $r = 0$, even if the metric periodicity is tuned to remove the conical singularity. 

Just as in the previous case, due to the singularity of the Weyl factor, the BTZ and conical geometries comprise one half of the boundary metric. By considering also the transformation $\bar{y} = \ell_3/r$ for $\bar{y} > 0$, a second identical copy is obtained. We illustrate the relevant two-dimensional sectors of these metrics in the right two panels of Figure~\ref{fig:two-copies-plots}.

\section{C-metric and brane configurations}
\label{sec:Cmetric}

We now turn to the C-metric and the possible pure tension branes in it, generalising the discussion in \cite{Hennigar:2026mxv} to include electric and magnetic charges. We consider the dyonic AdS C-metric, with metric
\begin{equation}
\label{ch_cmetric_xy}
\mathrm{d}s^2
=
\frac{1}{A^2(x-y)^2}
\left[
H(y)\mathrm{d}t^2
-\frac{\mathrm{d}y^2}{H(y)}
+\frac{\mathrm{d}x^2}{G(x)}
+G(x)\mathrm{d}\phi^2
\right],
\end{equation}
and gauge field
\begin{equation}
\label{gauge_field}
F
=
\frac{2}{A\ell_\star}
\left(
e \mathrm{d}y \wedge \mathrm{d}t
+g \mathrm{d} x \wedge \mathrm{d}\phi
\right) .
\end{equation}
The metric functions read
\begin{equation}
\label{cmetric_structure_functions}
H(y)
=
-\lambda-k y^2-\mu y^3-q^2y^4,
\qquad
G(x)
=
1-k x^2-\mu x^3-q^2x^4,
\qquad
q^2=e^2+g^2 , 
\end{equation}
where $e$ and $g$ are the dimensionless electric and magnetic
charge parameters. The parameters $A$ and $\lambda$ are related to the bulk cosmological length scale by
\begin{equation}
\ell_4
=
\frac{1}{A\sqrt{1+\lambda}} \, .
\end{equation}
The coordinates and the parameters
$\lambda,k,\mu,e,g$ are dimensionless, while $A$ has dimensions
of inverse length.

As in the uncharged AdS C-metric, the $\lambda > 0$ solution enjoys a symmetry under double analytic continuation. For $\lambda>0$, the transformation
\begin{equation}
    t\longrightarrow i\Phi,
    \qquad
    \phi\longrightarrow iT,
    \qquad
    e\longrightarrow-ie,
    \qquad
    g\longrightarrow-ig
\end{equation}
maps the solution to another member of the charged C-metric
family. After defining
\begin{equation}
    \label{DAC_coord_transform}
    x=-\sqrt{\lambda}\,\hat y,
    \qquad
    y=-\sqrt{\lambda}\,\hat x,
    \qquad
    T=\frac{\hat t}{\sqrt{\lambda}},
    \qquad
    \Phi=\frac{\hat\phi}{\sqrt{\lambda}},
\end{equation}
and
\begin{equation}
    \label{DAC_param_transform}
    \hat A=A\sqrt{\lambda},
    \qquad
    \hat\lambda=\frac{1}{\lambda},
    \qquad
    \hat k=-k,
    \qquad
    \hat\mu=\mu\sqrt{\lambda},
    \qquad
    \hat e=-\sqrt{\lambda}\,g,
    \qquad
    \hat g=-\sqrt{\lambda}\,e,
\end{equation}
the metric and gauge potential recover the forms
\eqref{ch_cmetric_xy} and \eqref{gauge_field} in terms of the hatted
coordinates and parameters. In particular,
\begin{equation}
    \hat q^2
    =
    \hat e^2+\hat g^2
    =
    \lambda q^2.
\end{equation}
Thus, the continuation exchanges $x$ and $y$, reverses the sign of $k$,  and
interchanges the electric and magnetic charge parameters, up to the
rescaling by $\lambda$.

\subsection{Charged C-metric root structures}

\begin{figure}[ht]
    \centering

    \newcommand{\rootpanel}[2]{%
        \begin{subfigure}[t]{0.31\textwidth}
            \centering
            \includegraphics[width=\linewidth]{#1}

            \parbox[c][2.2\baselineskip][c]{\linewidth}{%
                \centering
                \scriptsize #2\par
            }
        \end{subfigure}%
    }

    \newcommand{\rootgroup}[1]{%
        \par\vspace{3pt}
        \noindent
        \colorbox{black!12}{%
            \parbox{\dimexpr\linewidth-2\fboxsep\relax}{%
                \centering
                \small\bfseries #1%
            }%
        }%
        \par\vspace{2pt}
    }

    \rootgroup{$k=-1$}
    \vspace{5 pt}

    \rootpanel
        {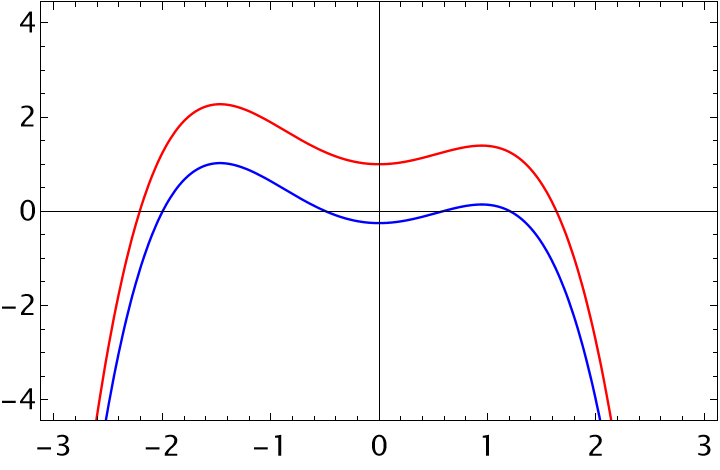}
        {$0<\lambda<G(\xi_+)-1$}
    \hfill
    \rootpanel
        {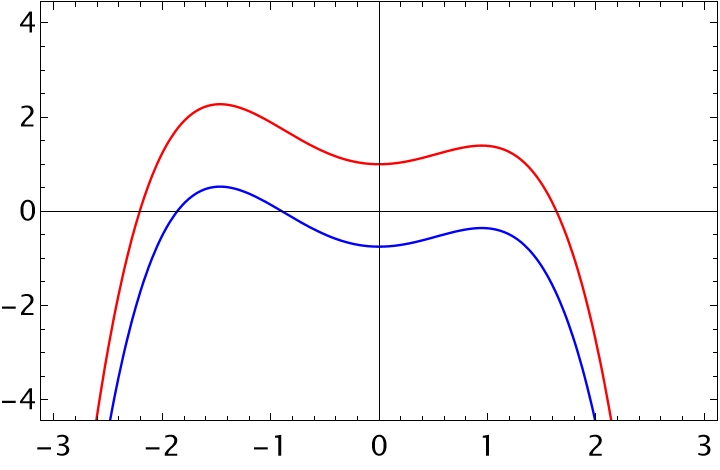}
        {\shortstack{
            $G(\xi_+)-1<\lambda$\\
            ${}<G(\xi_-)-1$
        }}
    \hfill
    \rootpanel
        {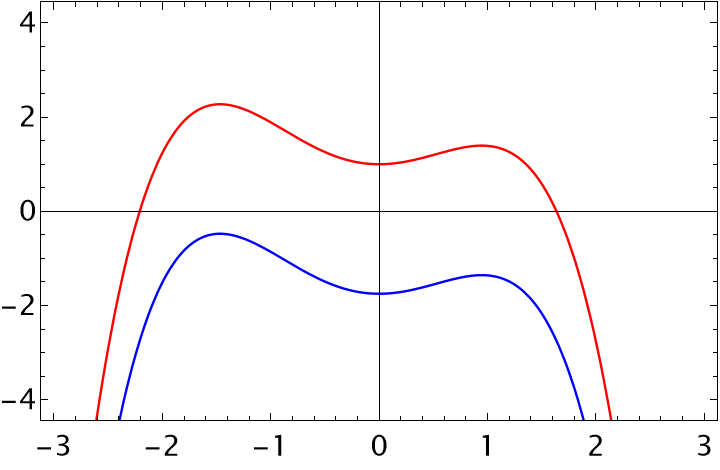}
        {$\lambda>G(\xi_-)-1$}

    \rootgroup{$k=1$, \quad $q^2<1/12$}
    \vspace{5 pt}

    \rootpanel
        {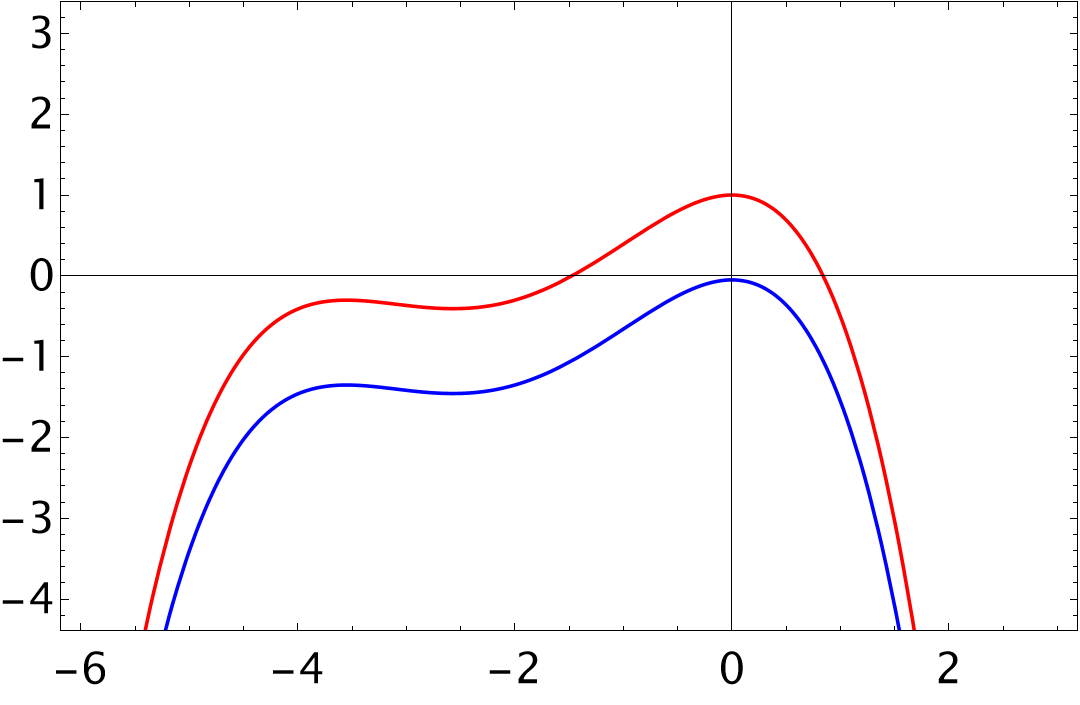}
        {\shortstack{
            $9\mu^2\leq32q^2$\\
            or $G(\xi_-)<0$
        }}
    \hfill
    \rootpanel
        {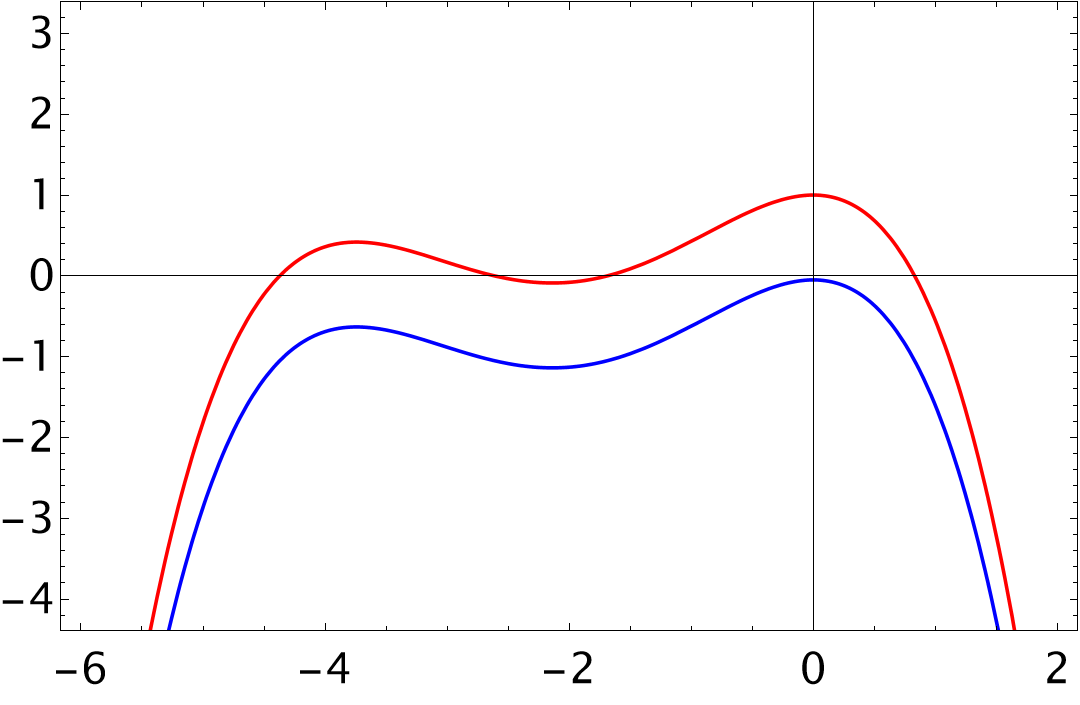}
        {$G(\xi_+)<0<G(\xi_-)<1$}
    \hfill
    \rootpanel
        {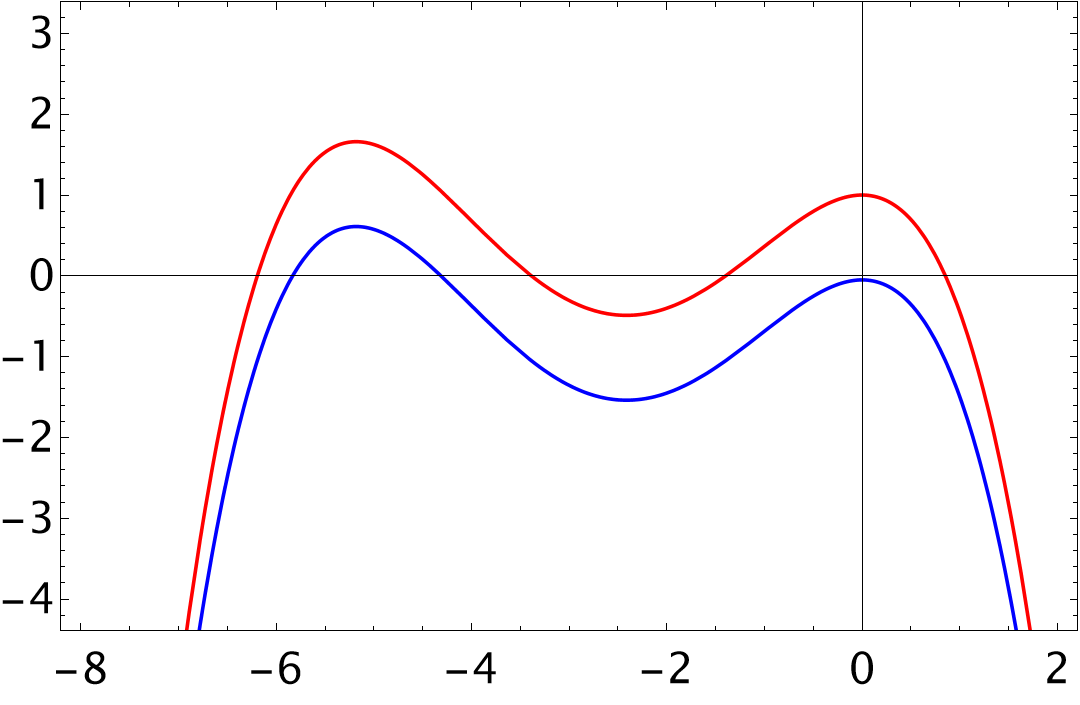}
        {\shortstack{
            $G(\xi_+)<0<1+\lambda$\\
            ${}<G(\xi_-)$
        }}

    \par\vspace{1pt}

    \rootpanel
        {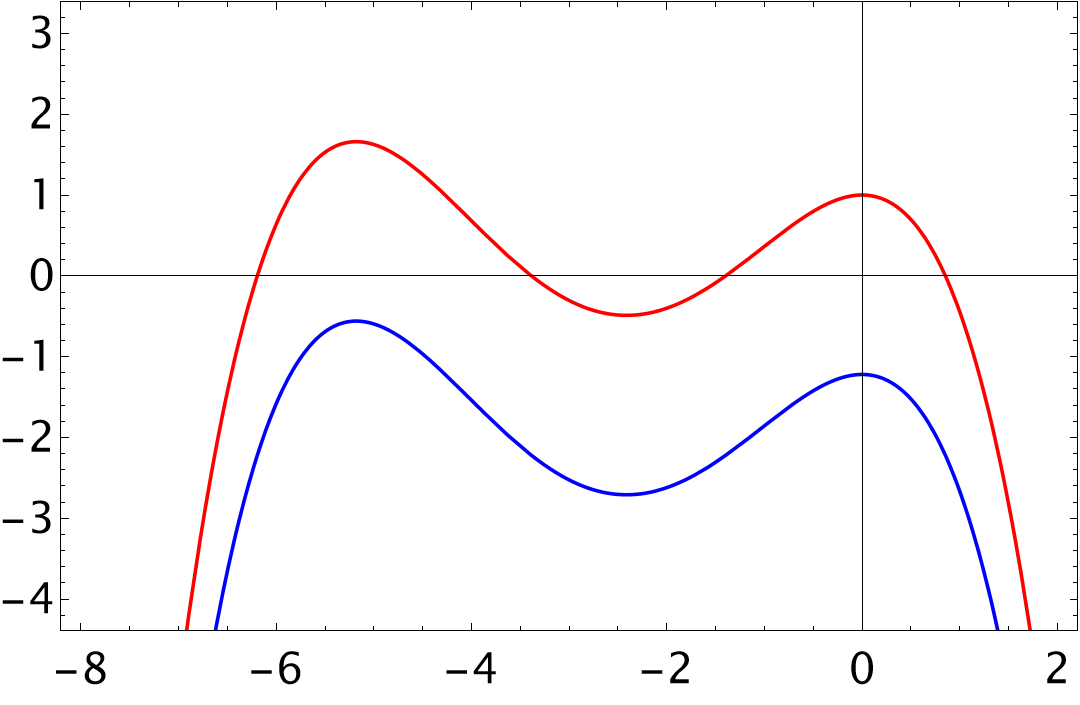}
        {\shortstack{
            $G(\xi_+)<0<G(\xi_-)$\\
            $G(\xi_-)<1+\lambda$
        }}
    \hfill
    \rootpanel
        {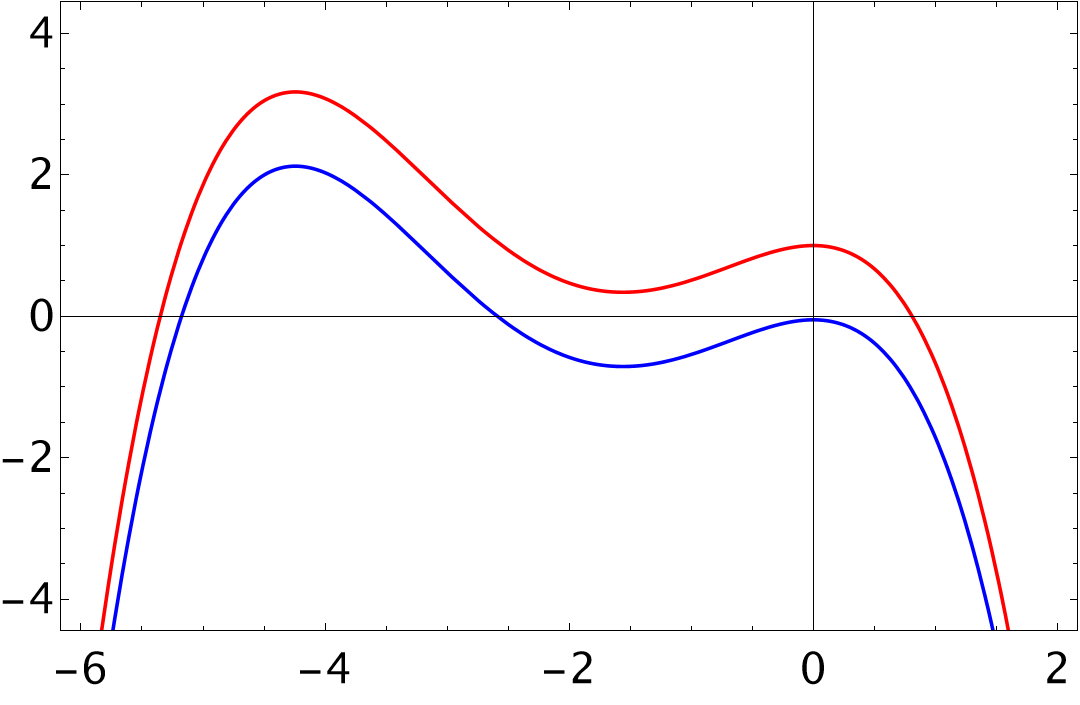}
        {\shortstack{
            $0<G(\xi_+)<1+\lambda$\\
            ${}<G(\xi_-)$
        }}
    \hfill
    \rootpanel
        {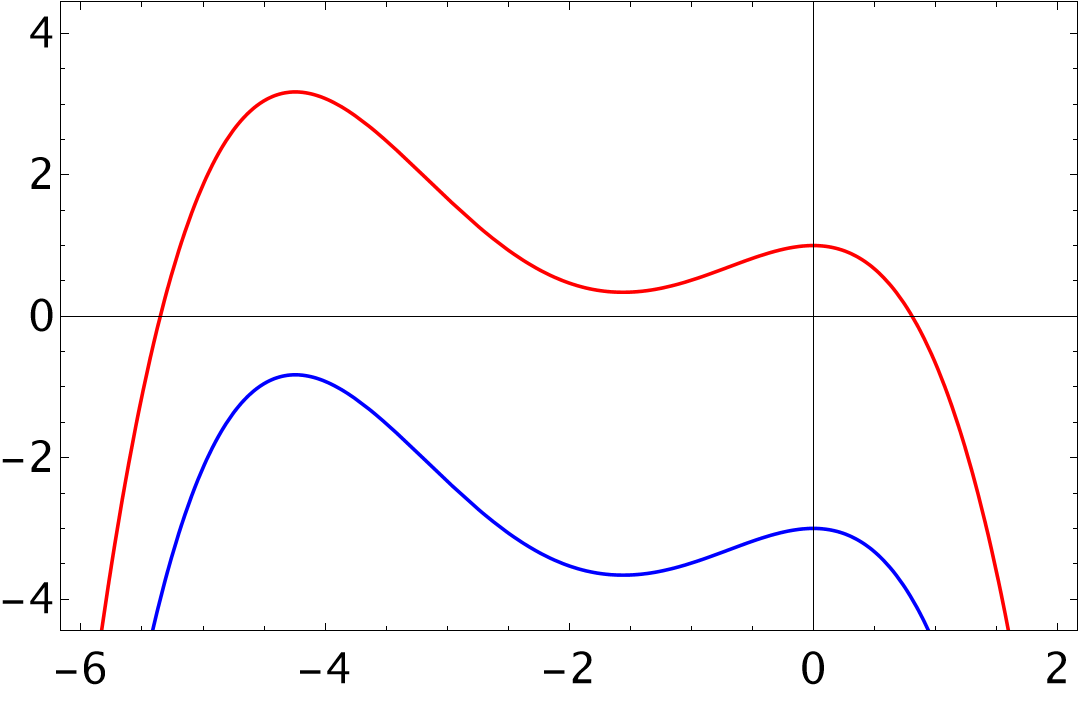}
        {\shortstack{
            $0<G(\xi_+)<G(\xi_-)$\\
            ${}<1+\lambda$
        }}

    \rootgroup{$k=1$, \quad $q^2>1/12$}
    \vspace{5 pt}
    
    \rootpanel
        {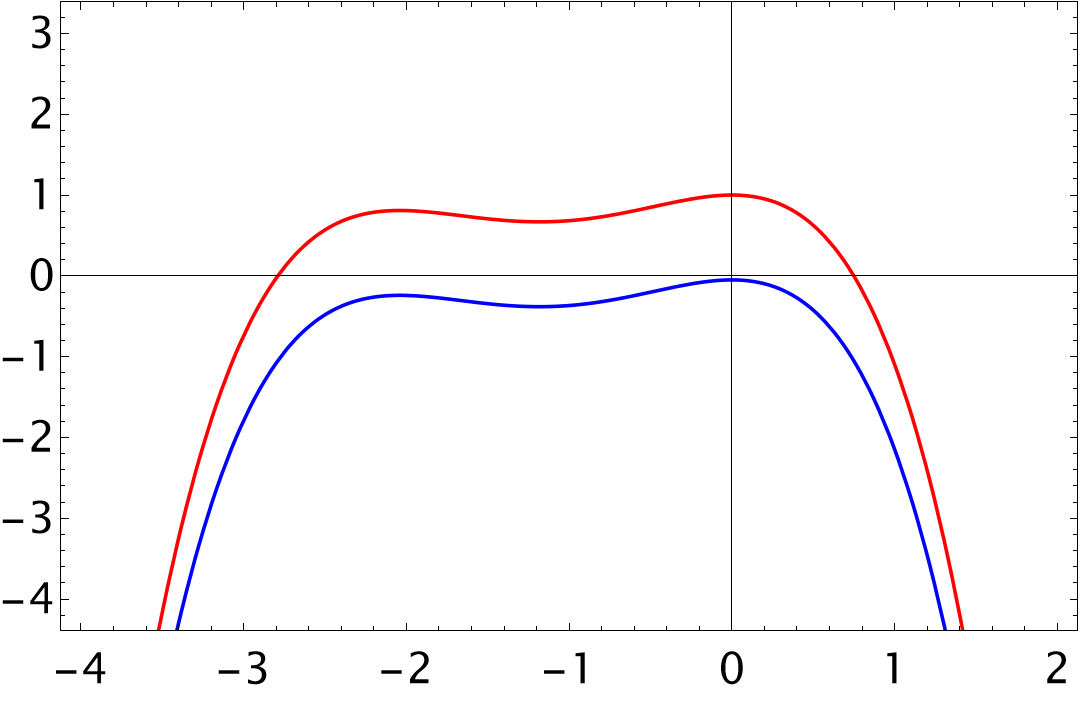}
        {$\mu<2q$}
    \hfill
    \rootpanel
        {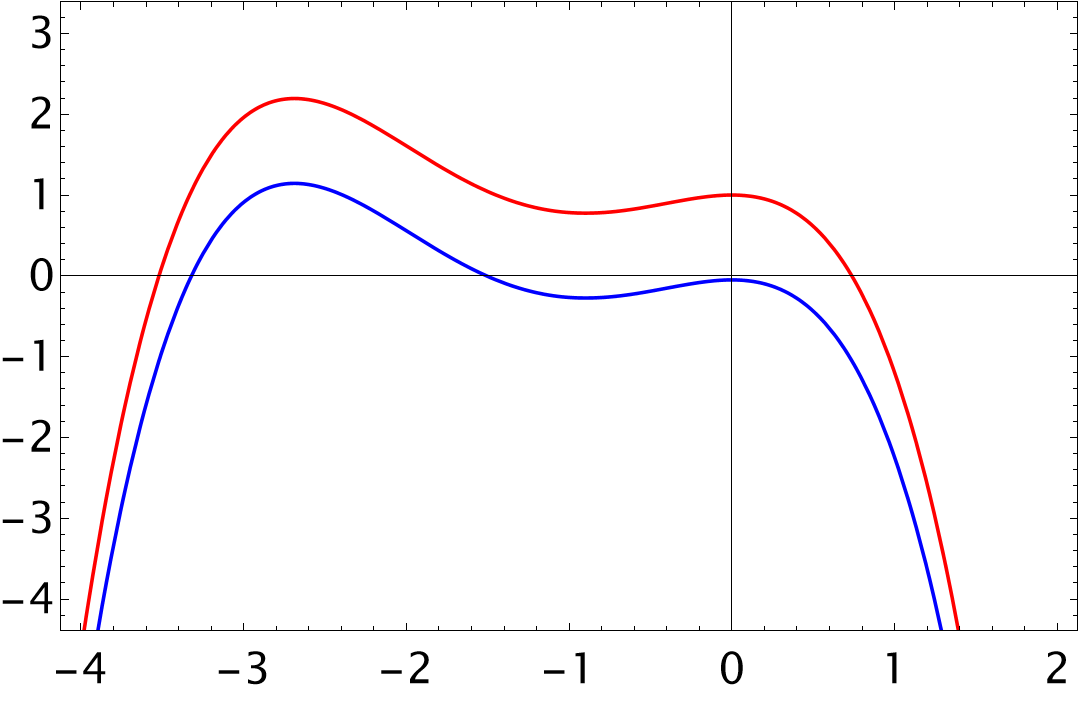}
        {\shortstack{
            $\mu>2q$\\
            $0<\lambda<G(\xi_-)-1$
        }}
    \hfill
    \rootpanel
        {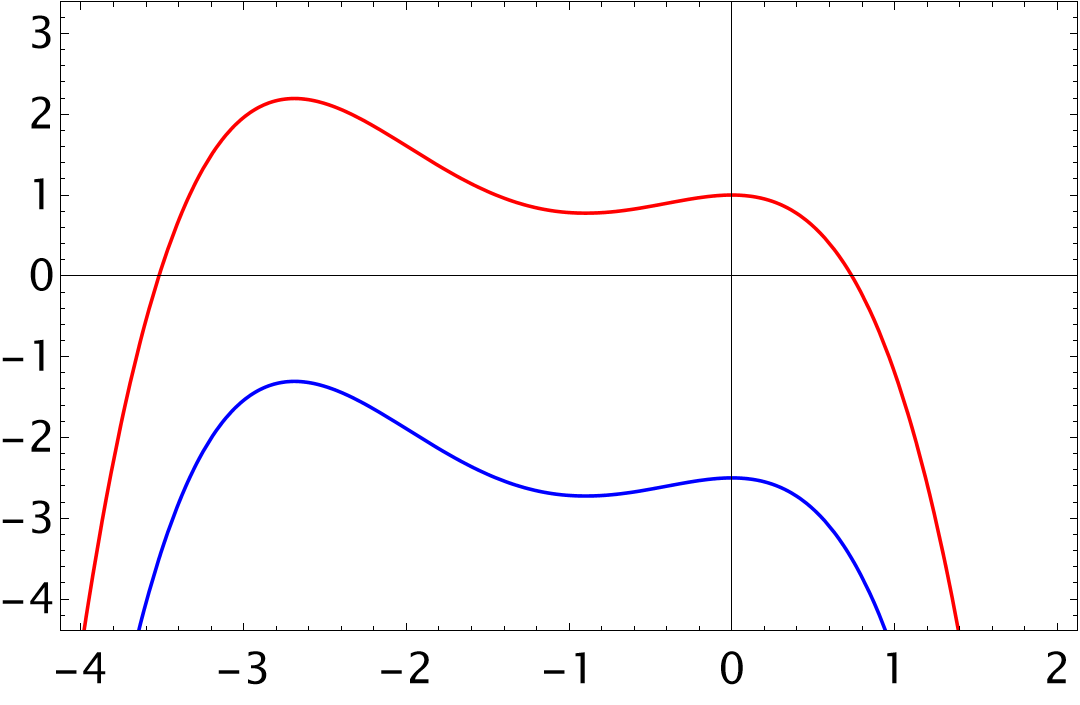}
        {\shortstack{
            $\mu>2q$\\
            $\lambda>G(\xi_-)-1$
        }}

    \caption{
        Representative root structures of $G(\xi)$ (red curves) and $H(\xi)$ (blue curves). 
    }
    \label{fig:charged-root-structures}
\end{figure}

The structure of the C-metric is determined by the roots of the functions $G(\xi)=1-k\xi^2-\mu\xi^3-q^2\xi^4$ and $H(\xi)=G(\xi)-(1+\lambda)$. Thus, the roots of $H$ are the intersections of the graph of $G$ with the horizontal line $G=1+\lambda$. We restrict throughout to generic parameter values, with $\lambda>0$ and $q^2>0$, excluding for brevity cases in which roots or stationary points become degenerate.\footnote{The C-metric makes sense for $\lambda <0$ but the brane constructions to follow restrict us to $\lambda >0$.} (These cases are achieved when our strict inequalities become equalities.) 
The functions $G$ and $H$ have the same stationary points. We denote these as $\xi_0=0$ and, when real,
\begin{equation}
\xi_\pm
=
\frac{-3\mu\pm\sqrt{9\mu^2-32kq^2}}
     {8q^2}.
\end{equation}

We will first discuss in detail the case with $\mu > 0$. Note that in the presence of charge, unlike the uncharged case discussed in \cite{Hennigar:2026mxv}, there can be physically interesting solutions with $\mu <0$. There is a transformation of the C-metric relating $\mu < 0$ to $\mu >0$, so the features of the solutions with $\mu <0$ can be recovered from the $\mu >0$ analysis, and we will describe these briefly at the end. We now list the different possible behaviours for $\mu >0$.

\paragraph{Root structures for \texorpdfstring{$k=-1$}{k=-1}.}

The nonzero stationary points always exist and satisfy $\xi_-<0<\xi_+$. Both $\xi_-$ and $\xi_+$ are local maxima, while $\xi_0=0$ is a local minimum, with $G(\xi_-)>G(\xi_+)>G(0)=1$. Since $G(\xi)\to-\infty$ as $\xi\to\pm\infty$, $G$ always has exactly two real roots, one negative and one positive. The roots of $H$ are classified as follows:

\begin{itemize}
    \setlength{\itemsep}{0.35em}

    \item If $0<\lambda<G(\xi_+)-1$, then $H$ has four real roots, two negative and two positive.

    \item If $G(\xi_+)-1<\lambda<G(\xi_-)-1$, then $H$ has two real roots, both negative.

    \item If $\lambda>G(\xi_-)-1$, then $H$ has no real roots.
\end{itemize}

\paragraph{Root structures for \texorpdfstring{$k=1$}{k=1}.}

The function $G$ is strictly decreasing for $\xi>0$, so it always has exactly one positive root. When $9\mu^2>32q^2$, the additional stationary points exist and satisfy $\xi_-<\xi_+<0$, with $\xi_-$ a local maximum and $\xi_+$ a local minimum. The possible root structures are:

\begin{itemize}
    \setlength{\itemsep}{0.6em}

    \item If $9\mu^2<32q^2$, then $\xi_0=0$ is the only stationary point and is the global maximum of $G$. In this case, $G$ has two real roots, one negative and one positive, while $H$ has no real roots for any $\lambda>0$.

    \item If $9\mu^2>32q^2$ and $\mu\leq 2q$, then $H$ has no real roots for any $\lambda>0$. If $G(\xi_-)>0>G(\xi_+)$, the function $G$ has four real roots, three negative and one positive; otherwise, it has two real roots, one negative and one positive.

    \item If $\mu>2q$, then $9\mu^2>32q^2$ automatically and $G(\xi_-)>1$. When $G(\xi_+)<0$, the function $G$ has four real roots, three negative and one positive; when $G(\xi_+)>0$, it has two real roots, one negative and one positive. Independently of this distinction, $H$ has two real roots, both negative, if $0<\lambda<G(\xi_-)-1$, and no real roots if $\lambda>G(\xi_-)-1$.
    \item Note the four-root configurations of $G$ can occur only when $q^2<1/12$. Consequently, if $q^2>1/12$, then $G$ has exactly two real roots for every value of $\mu$. For $k=1$, the complete condition for $H$ to possess real roots is
\begin{equation}
H\text{ has two real roots}
\quad\Longleftrightarrow\quad
\mu>2q
\quad\text{and}\quad
0<\lambda<G(\xi_-)-1.
\end{equation}
Whenever these roots exist, both are negative.
\end{itemize}

\begin{figure}[ht]
    \centering
    \setlength{\fboxsep}{2pt}

    \newcommand{\trianglepanel}[3]{%
        \begin{minipage}[t]{0.24\textwidth}
            \centering

            \colorbox{black!12}{%
                \parbox{\dimexpr\linewidth-2\fboxsep\relax}{%
                    \centering
                    \scriptsize\bfseries #1%
                }%
            }

            \vspace{1pt}

            \includegraphics[width=\linewidth]{#2}

            \parbox[c][2.3\baselineskip][c]{\linewidth}{%
                \centering
                \scriptsize #3\par
            }
        \end{minipage}%
    }


    \trianglepanel
        {$k=-1$}
        {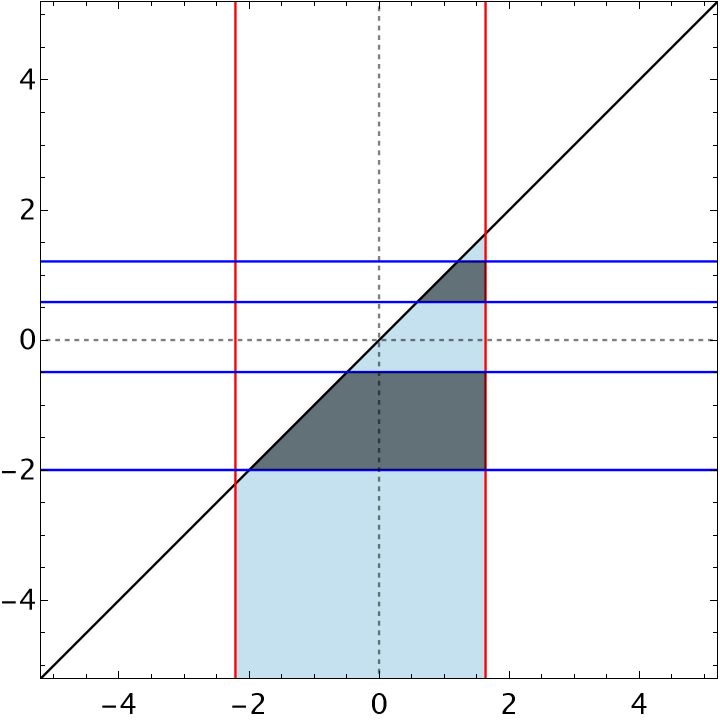}
        {$0<\lambda<G(\xi_+)$}
    \hfill
    \trianglepanel
        {$k=-1$}
        {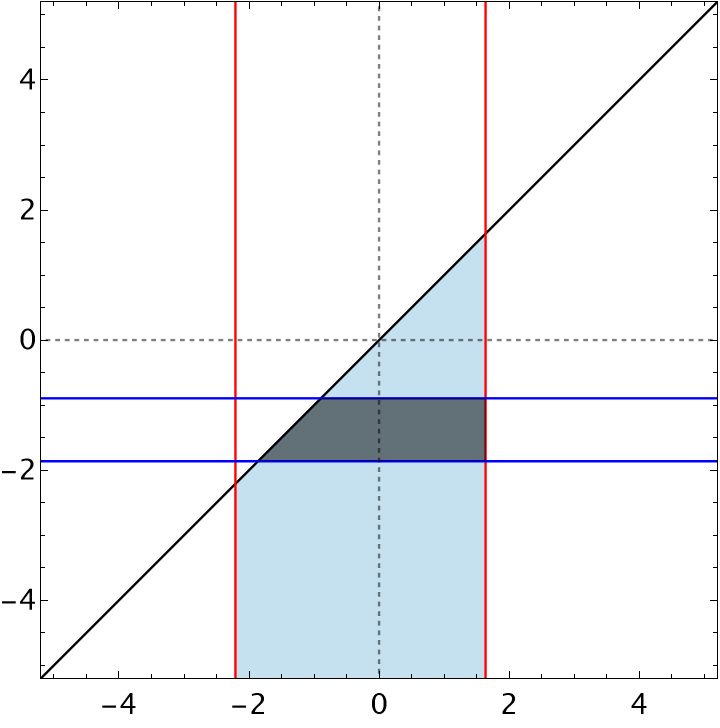}
        {\shortstack{
            $G(\xi_+)<\lambda$\\
            ${}<G(\xi_-)$
        }}
    \hfill
    \trianglepanel
        {$k=-1$}
        {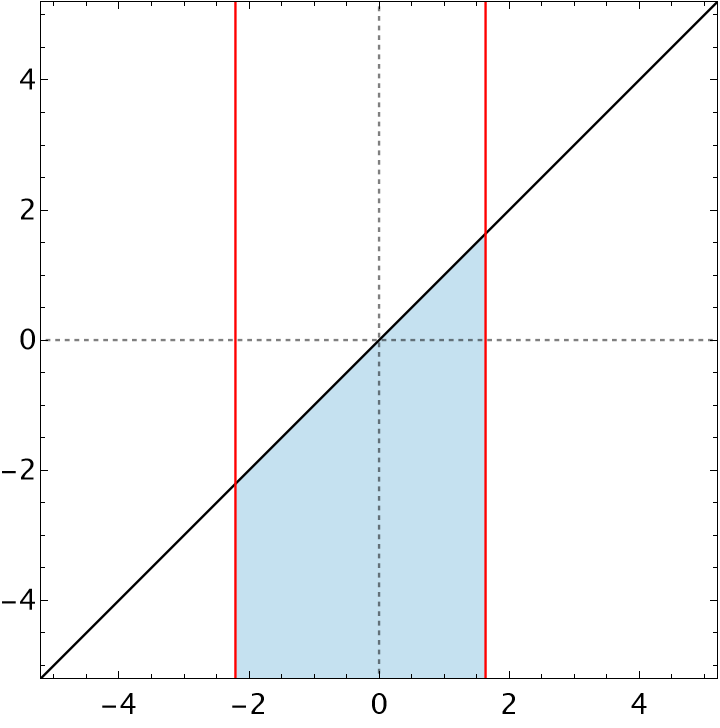}
        {$\lambda>G(\xi_-)$}
    \hfill
    \trianglepanel
        {$k=1$, \ $q^2<1/12$}
        {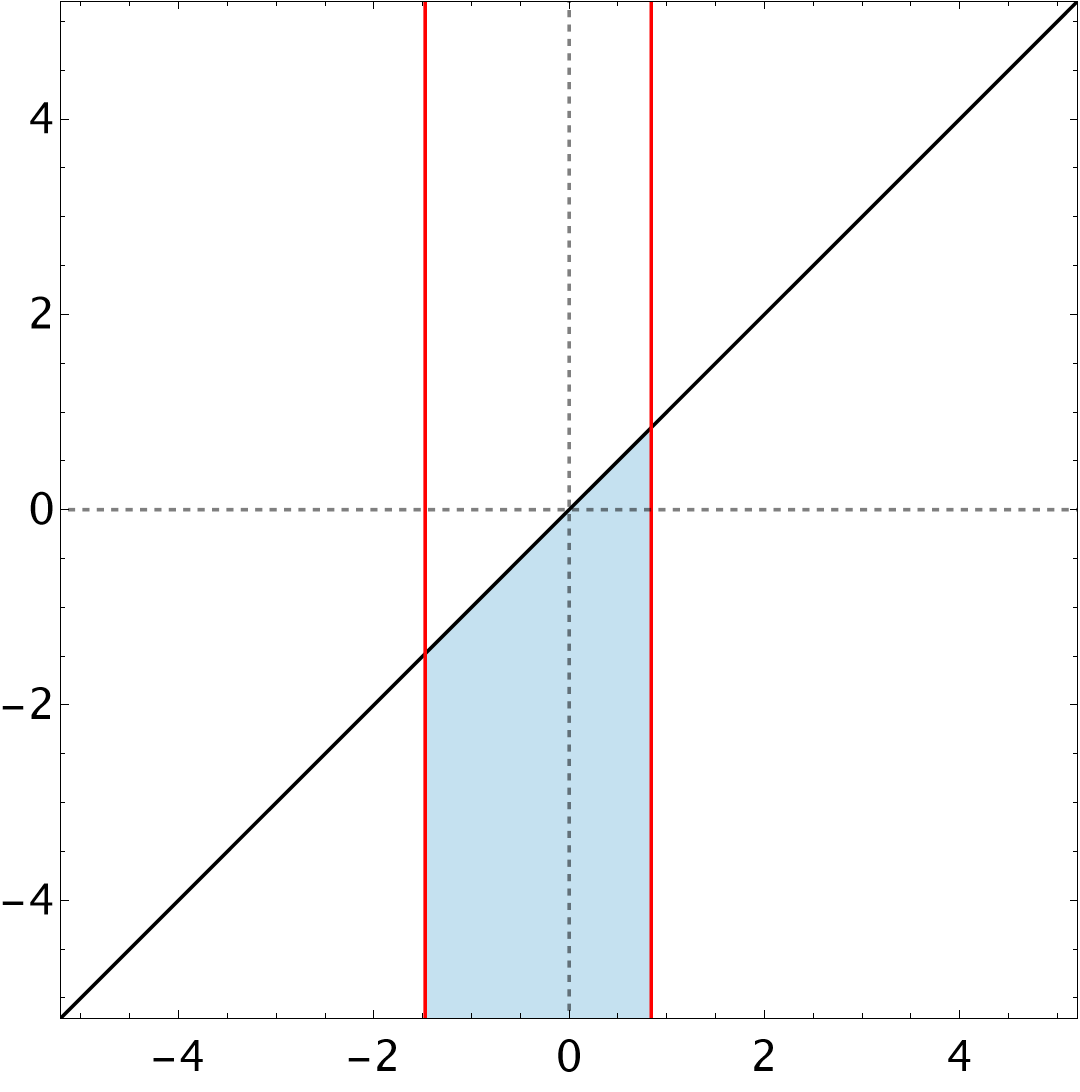}
        {\shortstack{
            $9\mu^2\leq32q^2$\\
            or $G(\xi_-)<0$
        }}

    \par\vspace{2pt}


    \trianglepanel
        {$k=1$, \ $q^2<1/12$}
        {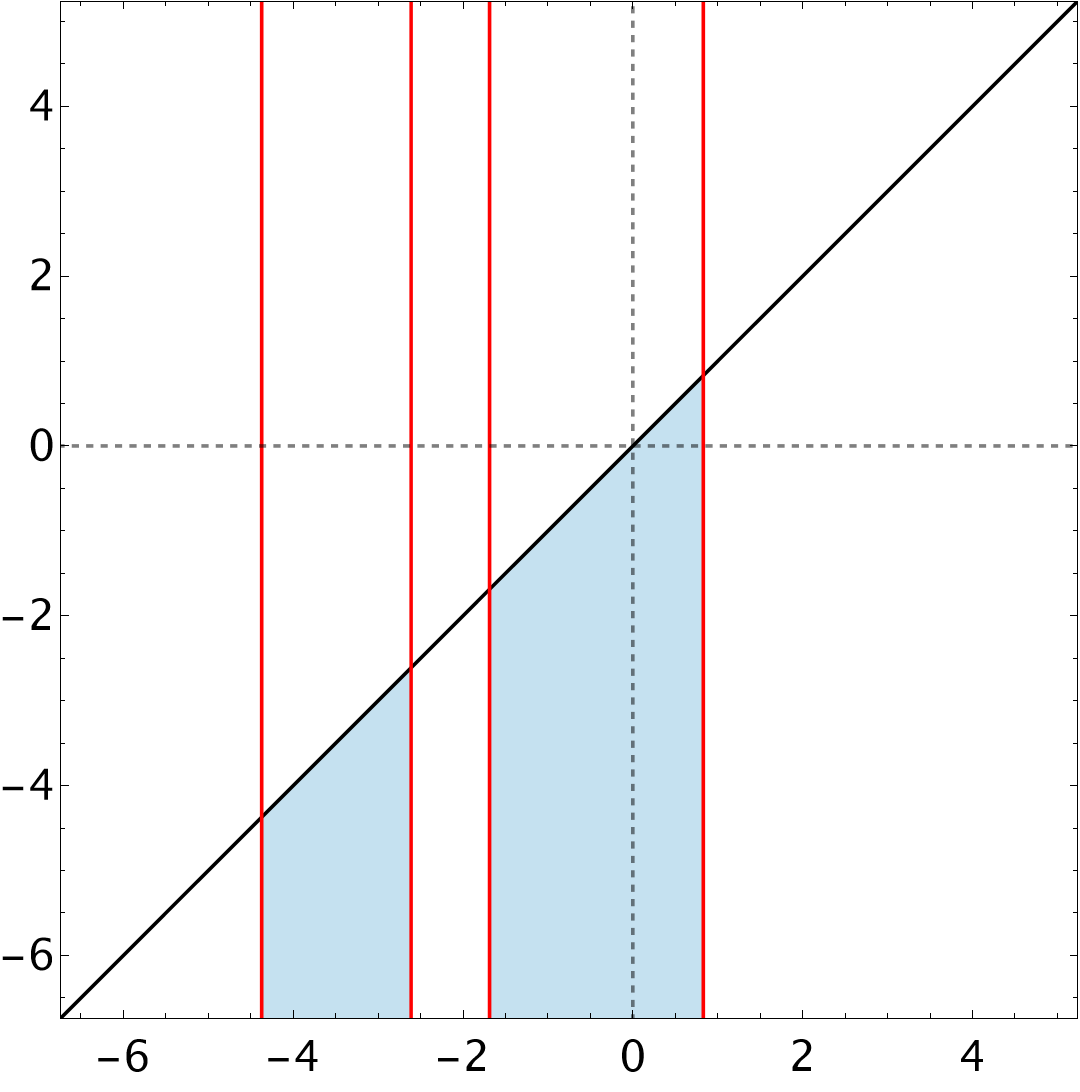}
        {$G(\xi_+)<0<G(\xi_-)<1$}
    \hfill
    \trianglepanel
        {$k=1$, \ $q^2<1/12$}
        {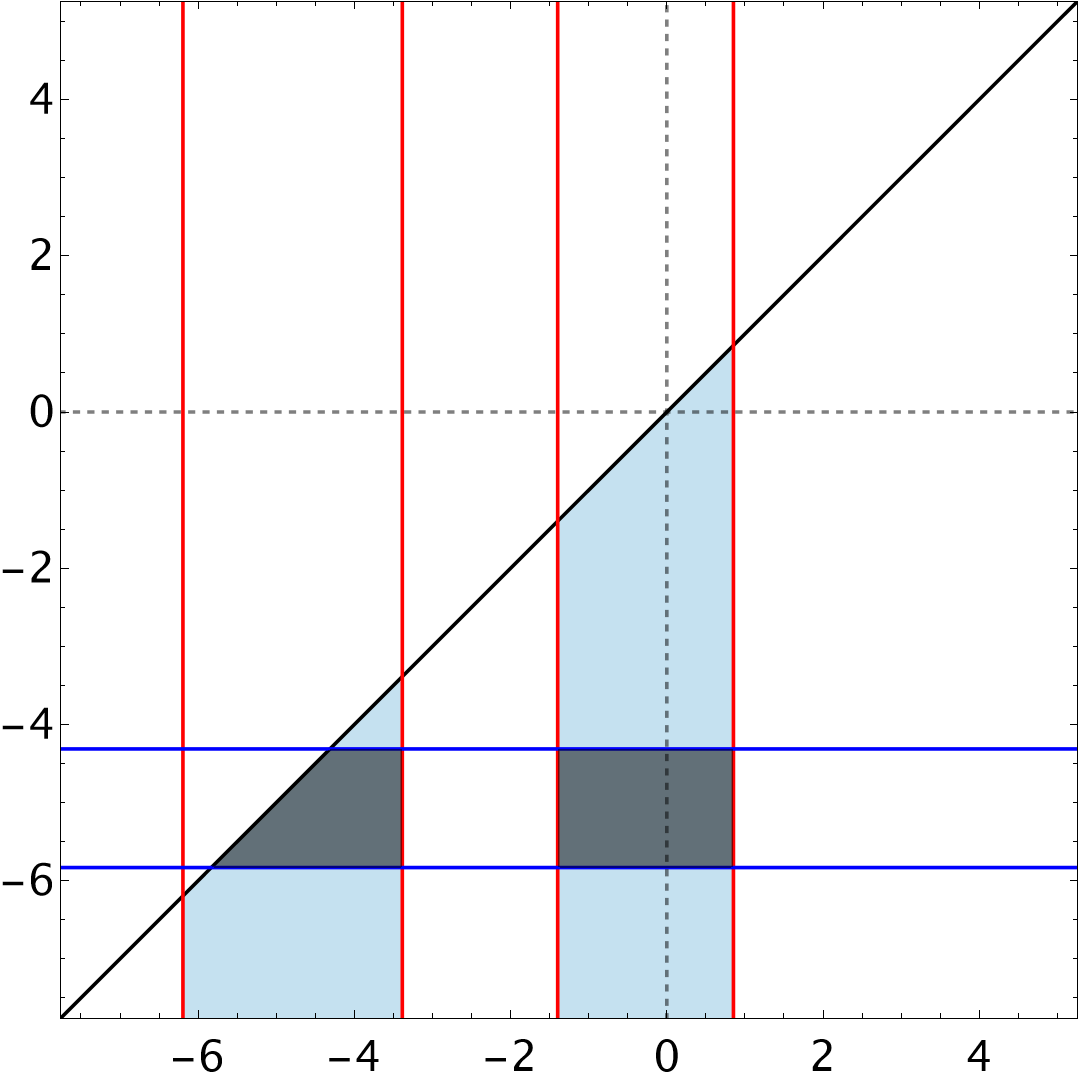}
        {\shortstack{
            $G(\xi_+)<0<1+\lambda$\\
            ${}<G(\xi_-)$
        }}
    \hfill
    \trianglepanel
        {$k=1$, \ $q^2<1/12$}
        {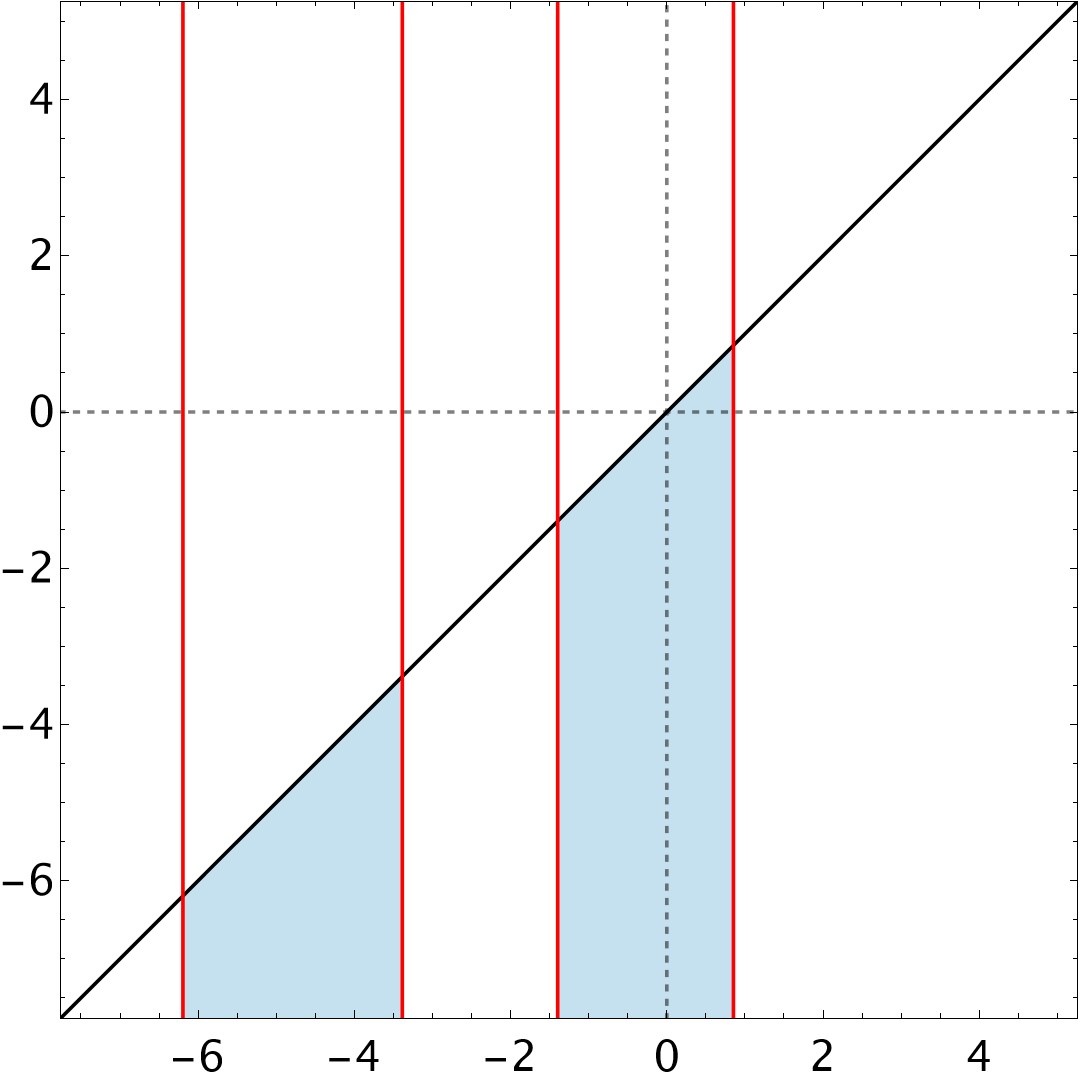}
        {\shortstack{
            $G(\xi_+)<0<G(\xi_-)$\\
            ${}<1+\lambda$
        }}
    \hfill
    \trianglepanel
        {$k=1$, \ $q^2<1/12$}
        {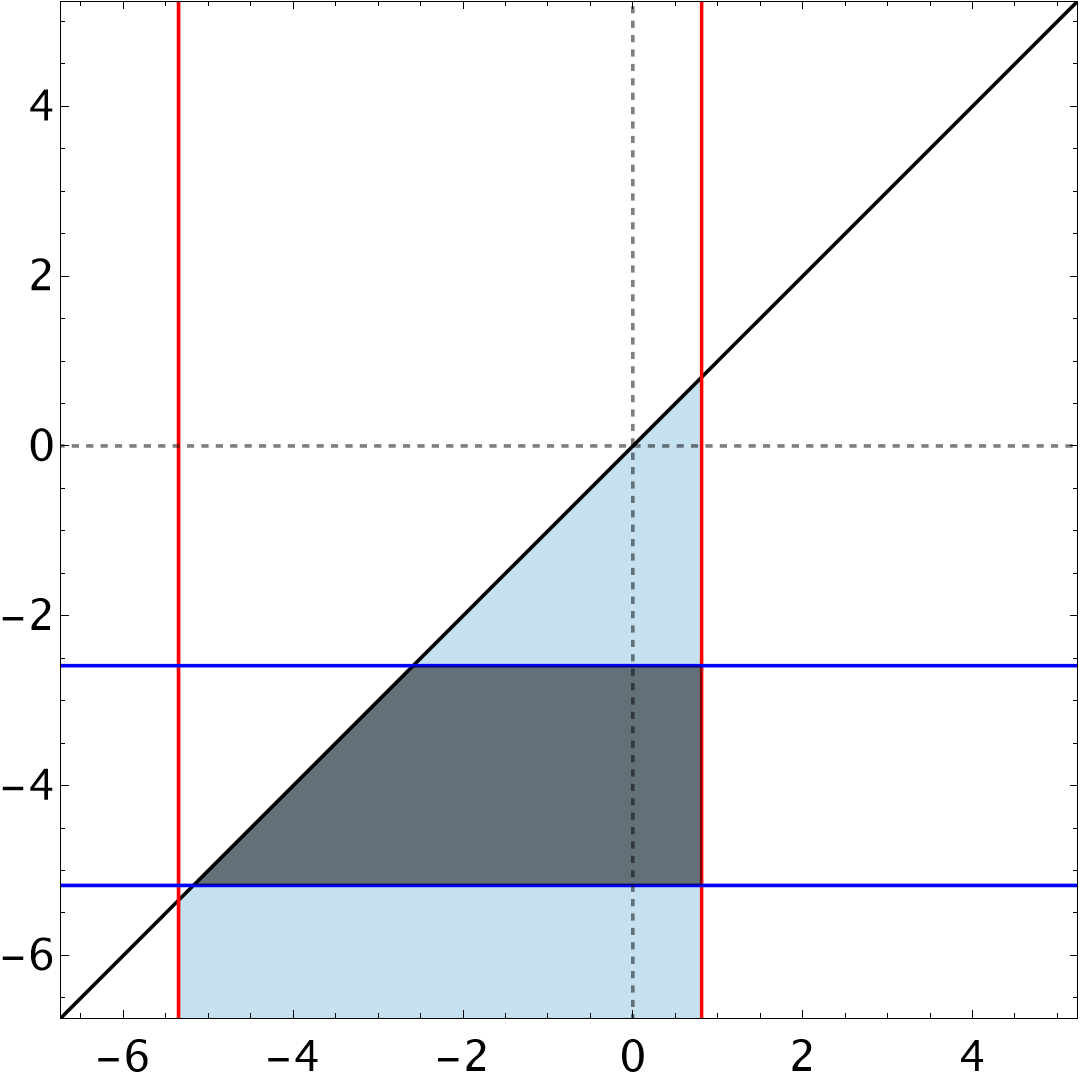}
        {\shortstack{
            $0<G(\xi_+)<1+\lambda$\\
            ${}<G(\xi_-)$
        }}

    \par\vspace{2pt}


    \trianglepanel
        {$k=1$, \ $q^2<1/12$}
        {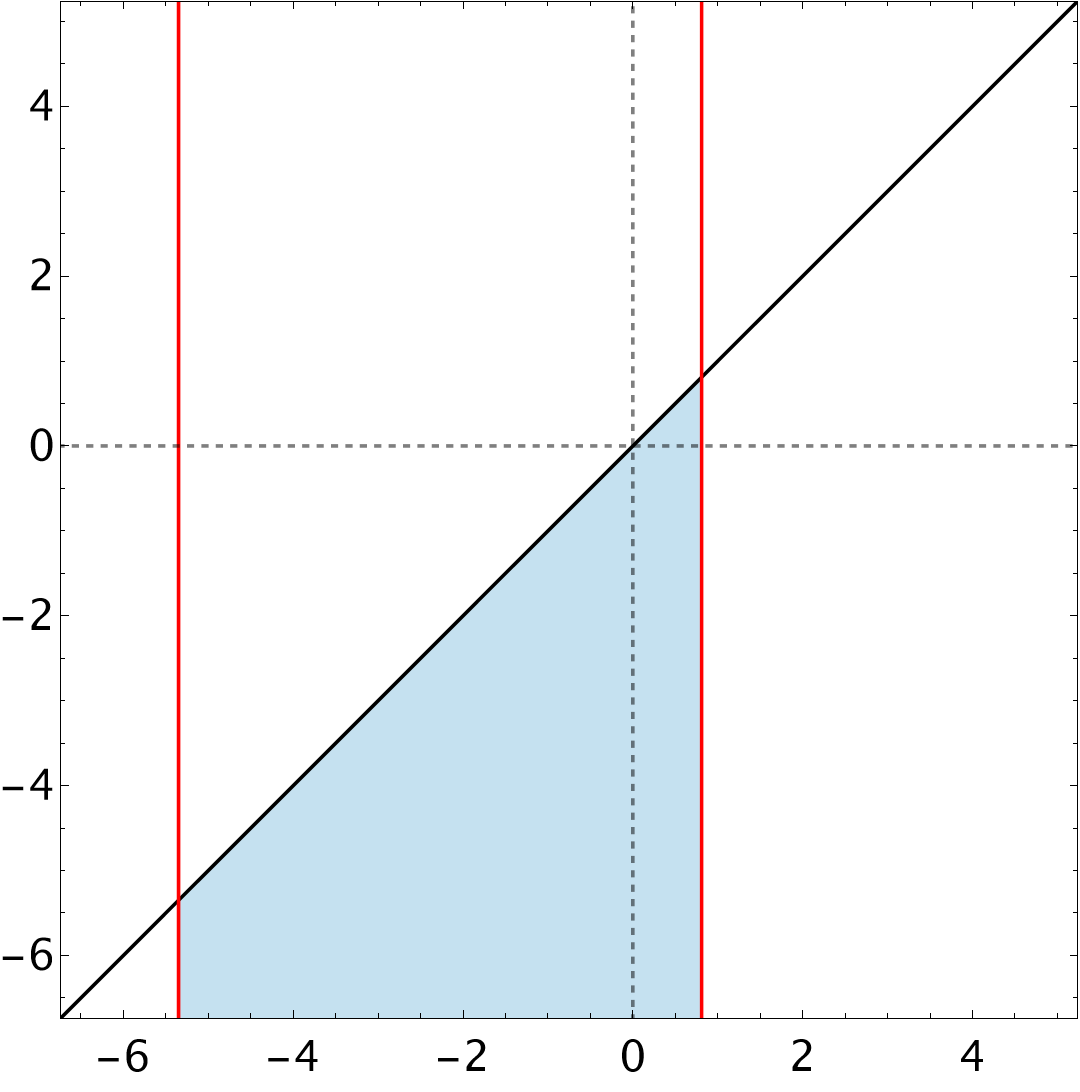}
        {\shortstack{
            $0<G(\xi_+)<G(\xi_-)$\\
            ${}<1+\lambda$
        }}
    \hfill
    \trianglepanel
        {$k=1$, \ $q^2>1/12$}
        {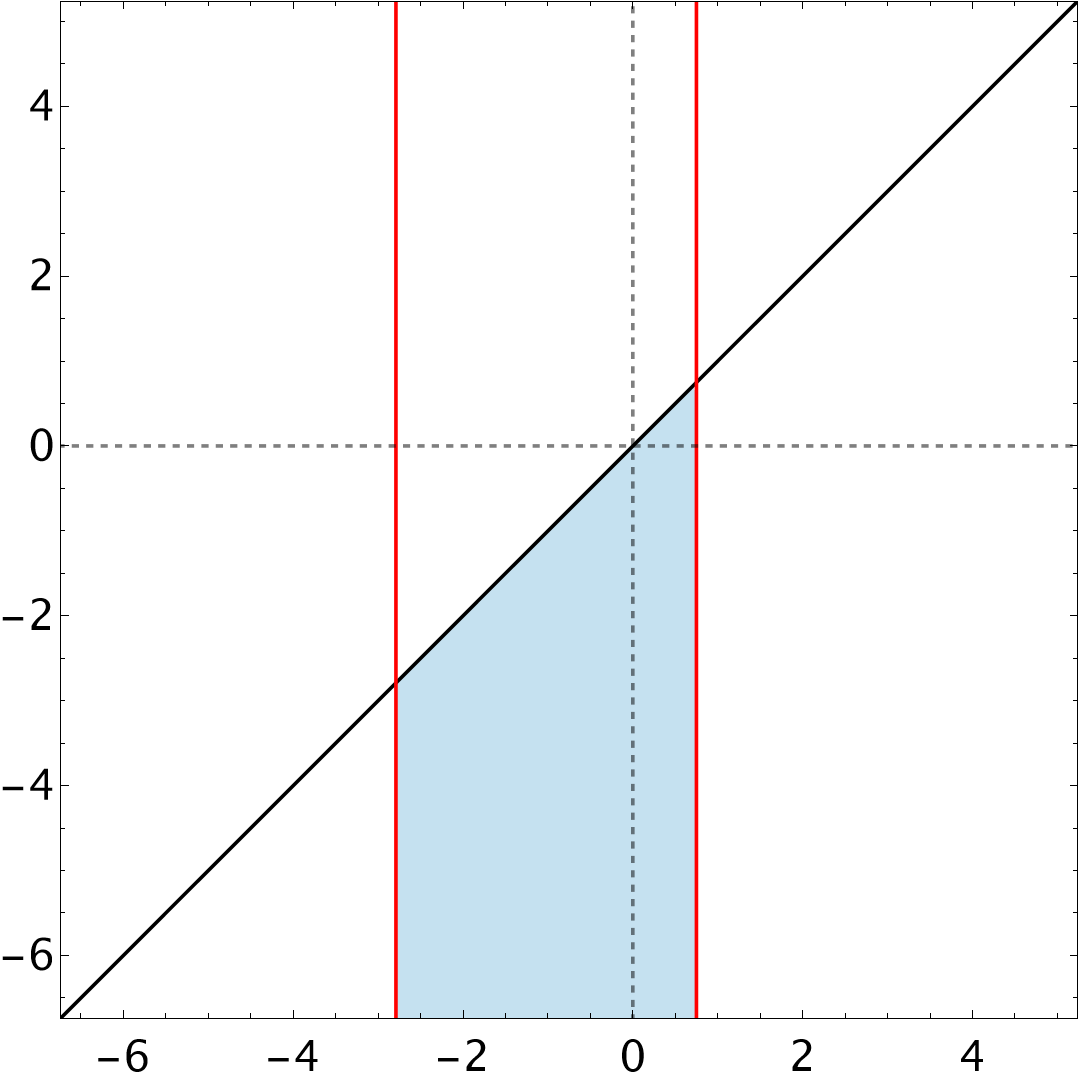}
        {$\mu<2q$}
    \hfill
    \trianglepanel
        {$k=1$, \ $q^2>1/12$}
        {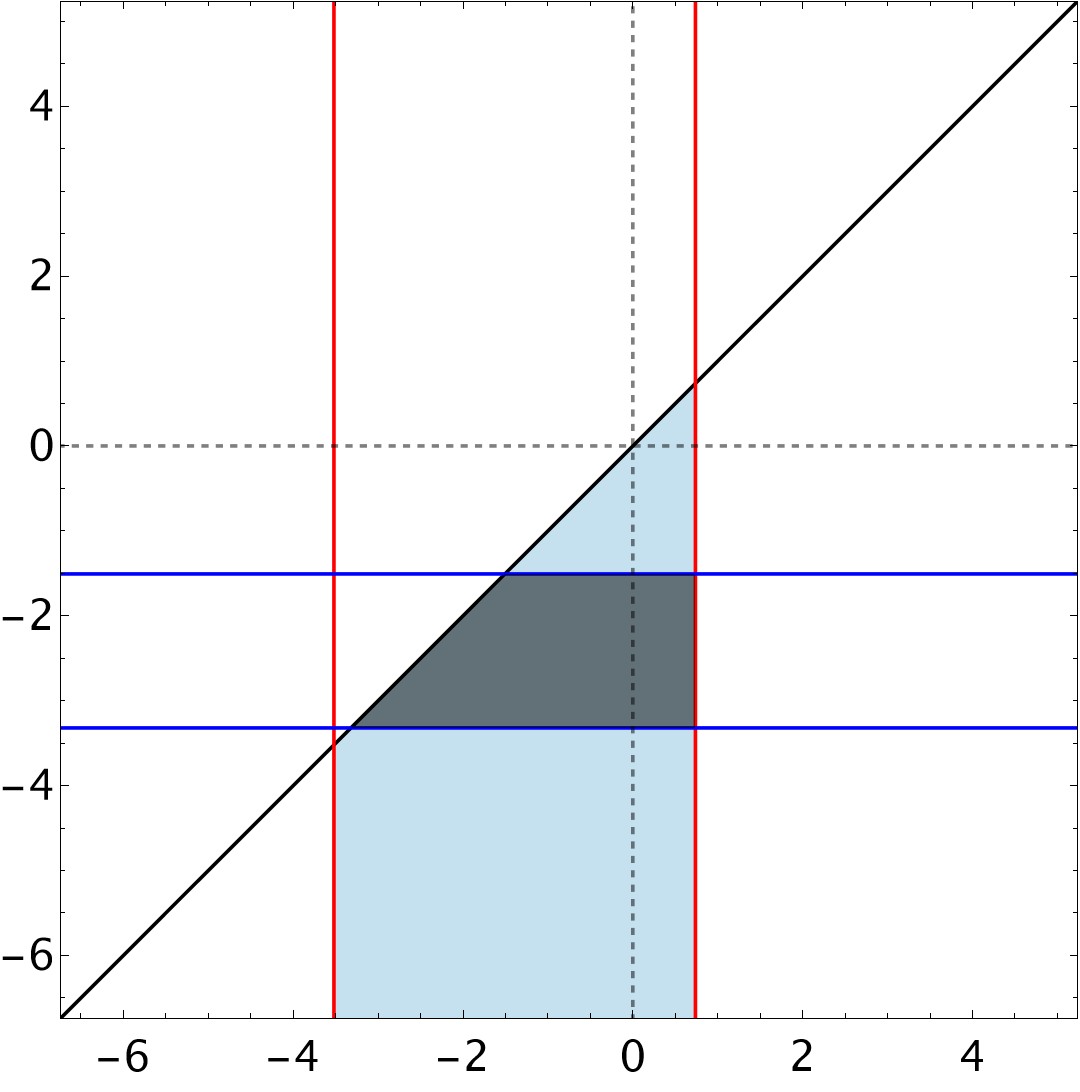}
        {\shortstack{
            $\mu>2q$\\
            $0<\lambda<G(\xi_-)-1$
        }}
    \hfill
    \trianglepanel
        {$k=1$, \ $q^2>1/12$}
        {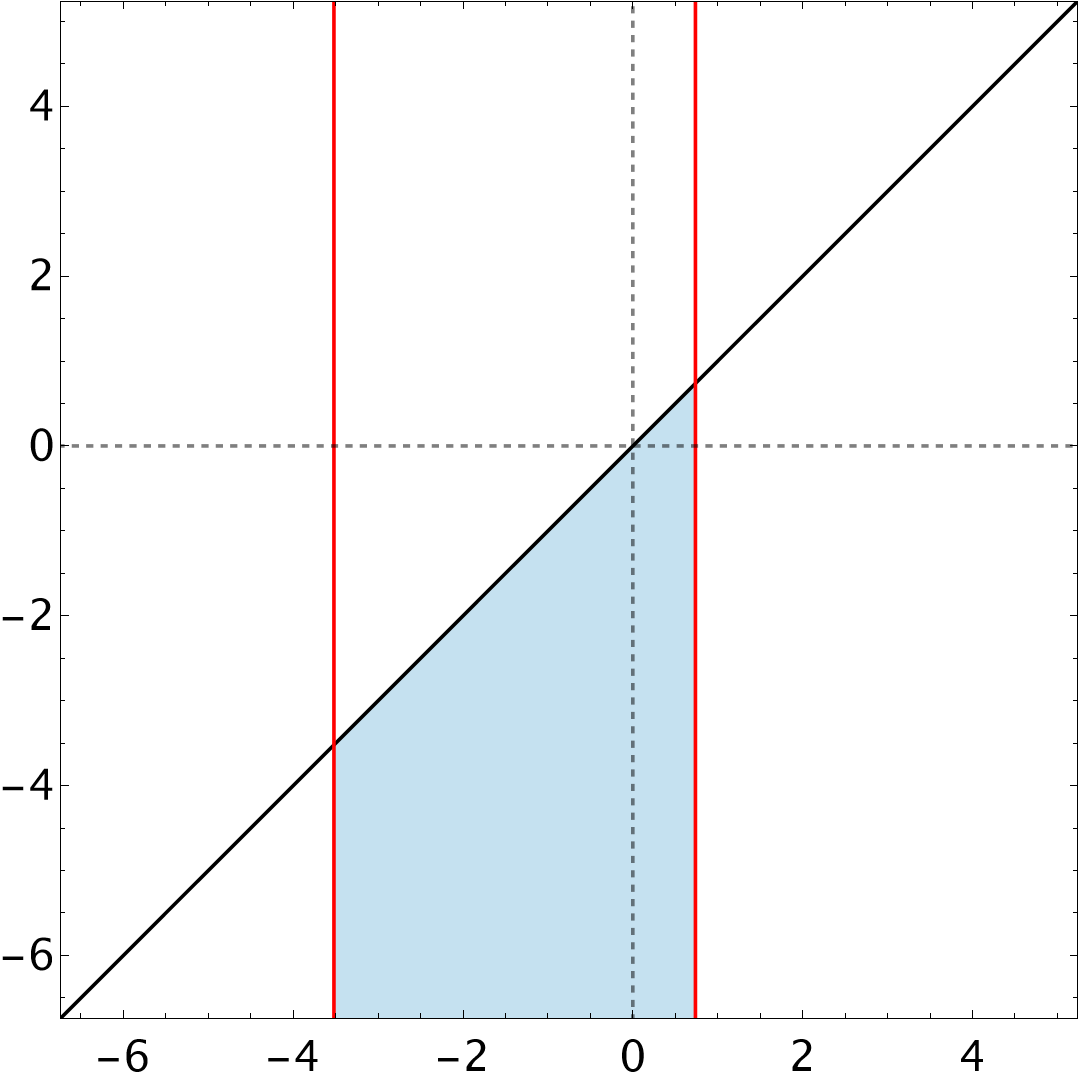}
        {\shortstack{
            $\mu>2q$\\
            $\lambda>G(\xi_-)-1$
        }}

    \caption{
        Parameter space in $(x,y)$ coordinates. Here the solid, black line is the asymptotic boundary $y = x$. Dashed gray lines are possible locations of branes. The red lines are zeroes of $G(x)$ while the blue lines are zeros of $H(y)$. Blue-shaded regions correspond to $H(y) < 0$ (untrapped surfaces), while gray-shaded regions correspond to $H(y) > 0$ (trapped surfaces). White regions are not relevant, either because they lie in the region $y > x$, or because $G(x) < 0$ there. 
    }
    \label{charged_triangle_plots}
\end{figure}

Characteristic behaviours of the functions $G$ and $H$ are displayed in Figure~\ref{fig:charged-root-structures}. 
Figure~\ref{charged_triangle_plots} shows $(x,y)$ parameter space plots for the same root configurations. In this plot, blue shaded regions correspond to untrapped surfaces where $H(y) < 0$ while gray-shaded regions correspond to trapped surfaces where $H(y) > 0$. Curvature singularities occur when $x, y \to \pm \infty$. By requiring Lorentzian geometries, we restrict ourselves to regions where $G(x) > 0$ --- regions where $x < y$ that are shaded in white correspond to $G(x) < 0$. In total there are 12 qualitatively different behaviours. 

Of these cases, only one corresponds to a compact event horizon, that with $k = 1$, $q^2 < 1/12$ and $G(\xi_+) < 0 < 1+\lambda < G(\xi_-)$. This would be the physically interesting case if we considered the C-metric on its own. When we consider the inclusion of branes, we can be interested in other cases; non-compact horizons can become compact on including an $x=0$ brane. We are however interested only in cases where $H(y)$ has negative roots, so that the singularity is hidden by an event horizon. This event horizon lies at the largest negative root of $H(y)$, which we call $y=y_-$. If $H(y)$ also has positive roots, the smallest positive root will correspond to an acceleration horizon. 

\paragraph{The case of $\mu < 0$.}

So far our analysis of the possible root structures has been restricted to $\mu > 0$. The case of $\mu < 0$ can be understood by symmetry. Writing
$G_\mu$ and $H_\mu$ to display their dependence on $\mu$, one has
\begin{equation}
    G_{-\mu}(\xi)=G_\mu(-\xi),
    \qquad
    H_{-\mu}(\xi)=H_\mu(-\xi),
    \qquad
    \xi_\pm(-\mu)=-\xi_\mp(\mu).
\end{equation}
At the level of the full C-metric, the transformation that maps the metric back into itself reads,
\begin{equation}
    (x,y,\mu,e,g)
    \longmapsto
    (-x,-y,-\mu,-e,-g) \, .
\end{equation}
Thus all root configurations for
$\mu<0$ are obtained by reflection through the origin. 
This reflection exchanges the regions
$x>y$ and $x<y$, so a chosen asymptotic region or brane interval must
be reflected as well. 

We obtain $\mu <0$ cases where $H$ has negative roots from $\mu >0$ cases where $H$ has positive roots. The only such case is $k=-1$, $0 < \lambda < G(\xi_+)-1$, so these are the physical solutions for $\mu <0$. The black hole solution \eqref{rnAdS} similarly has physical solutions with $\mu <0$ for $\kappa=-1$, but the uncharged C-metric does not, so this is a new feature in the charged case.  

\paragraph{Global considerations.}  The zeroes of $H(y)$ correspond to Killing horizons. In ingoing Eddington-Finkelstein coordinates, the horizon Killing vector field is $k^a=(\partial/\partial v)^a$, and at the roots of $H$, it becomes $k_a=(dy)_a$ which is normal to a null hypersurface. The roots of $G(x)$ are axes where the azimuthal coordinate $\phi$ shrinks to zero size. Regularity at an axis requires that $\phi$ is identified periodically $\phi \sim \phi + \Delta \phi$. Generically, we can only make the spacetime regular at one axis. We will choose to make it regular at the positive root $x=x_+$, by choosing 
\begin{equation}\label{phi_periodicity}
    \Delta\phi=\frac{4\pi}{|G'(x_+)|}=\frac{4\pi}{|2kx_++3\mu x_+^2+4q^2x_+^3|} \, .
\end{equation}
We can choose to make the gauge potential regular at the positive root $x=x_+$ and at the horizon $y=y_-$ by taking 
\begin{equation}
\mathcal A
=
\frac{2}{A\ell_\star}
\left(
e (y-y_-)  \wedge \mathrm{d}t
+g  (x-x_+) \wedge \mathrm{d}\phi
\right) .
\end{equation}

\FloatBarrier

\subsection{Branes in the charged C-metric}

So far, we have focused on the properties of the charged AdS$_4$ C-metric as a bulk solution. Our goal is to introduce branes into this spacetime and thereby construct solutions to a three-dimensional semiclassical theory of gravity. The construction of branes in the charged C-metric proceeds much as in the uncharged case considered in \cite{Hennigar:2026mxv}, but we give the details here to make the current treatment self-contained. 

Incorporating branes requires supplementing the bulk gravitational action with additional terms that enforce the appropriate junction conditions on the brane. In particular, we must account for the junction conditions satisfied by both the metric and the gauge field~\cite{Israel:1966rt, Lemos:2021jtm}. Let us consider the details for the metric first. The bulk region is split into two regions $\mathcal{M}_+$ and $\mathcal{M}_-$ which are glued together along a co-dimension 1 brane $\mathcal{B}$. The induced metric $h_{ij}$ must agree on the two sides, but its normal derivative need not.  Consequently, the gluing will generally produce a distributional source localized on $\mathcal{B}$.  We shall restrict ourselves to the case where this source is pure tension. 

Let $n_\pm^\mu$ denote the outward-pointing unit normal to $\mathcal{B}$ as viewed from $\mathcal{M}_\pm$, and define the corresponding extrinsic curvatures by
\be 
K_{ij}^{(\pm)} = h_i^\mu h_j^\nu \nabla_\mu n_\nu^{(\pm)} \, .
\ee
In addition to the bulk action, a well-posed variational problem requires a Gibbons-Hawking-York term on each side of the brane, together with an action for the localized distributional source of stress-energy on the brane. These contributions to the action take the form
\begin{align}
    I_{\rm tot} = I_{\rm bulk}^{(+)} + I_{\rm bulk}^{(-)} + \frac{1}{8 \pi G_4} \int_{\mathcal{B}} {\rm d}^3 x \sqrt{-h} \left(K^{(+)} + K^{(-)} \right) - \tau \int_{\mathcal{B}} {\rm d}^3 x \sqrt{-h} \, .
\end{align}
Variation of this action with respect to the brane metric $h_{ab}$ yields the Israel junction conditions. These can be written in the form\footnote{Our convention is that the normal vector is outward pointing with respect to $\mathcal{M}_\pm$. If instead we take a common orientation for the normal, this takes the perhaps more familiar form in terms of relative differences of the extrinsic curvature.}
\be 
\sum_{s=\pm} \left(K_{ij}^{(s)} - K^{(s)} h_{ij} \right) =  8 \pi G_4 S_{ij}
\ee
where $S_{ij}$ is the distributional stress tensor arising from the pure tension brane,
\be 
S_{ij} = - \tau h_{ij} \, .
\ee

If the brane is placed at an umbilic surface, then the Israel junction conditions can be satisfied. Consider a two-sided $\mathbb{Z}_2$ configuration for which we have
\be 
K^{(+)}_{ij} = K^{(-)}_{ij} = \frac{1}{\ell} h_{ij} \, .
\ee
The Israel junction conditions are then satisfied provided that 
\be \label{eq:umbilic_tension}
\tau = \frac{1}{2 \pi \ell G_4} \, .
\ee
In the C-metric, surfaces of constant $x$ or $y$ where $G'(x)$ or $H'(y)$ vanish respectively are umbilic surfaces. These are the locations we can put branes. We consider the possibilities in more detail shortly. 

Let us now consider the junction conditions for the Maxwell field. Let $\mathcal{A}_i$ be the gauge potential projected onto the brane. If the Maxwell flux across the brane does not match, then the brane action must also include a surface current, 
\be  \label{eq:branecurrent}
I_{\mathcal{B}}^{(j)} = \int_{\mathcal{B}} {\rm d}^3 x \sqrt{-h} \mathcal{A}_i j^i_{\mathcal{B}} \, .
\ee
The surface current is locally conserved $\nabla_i j^i_{\mathcal{B}} = 0$ due to gauge invariance. Define the flux normal to the boundary,
\be 
\mathcal{E}^i_{(\pm)} = n_\mu^{(\pm)} F_{(\pm)}^{\mu i} \, .
\ee
A well-posed variational principle then requires that
\be 
F_{(+)}^{ij} = F_{(-)}^{ij}\qquad \text{and}
\qquad \mathcal{E}_{(+)}^i + \mathcal{E}_{(-)}^i = g_\star^2 \, j^i_{\mathcal{B}} \, .
\ee

We shall impose that the full solution, both metric and gauge field, is $\mathbb{Z}_2$ symmetric across the brane. This then yields equal fluxes normal to the brane,
\be
\mathcal{E}_{(-)}^i = \mathcal{E}_{(+)}^i \, .
\ee
which add to produce a current localized on the brane, 
\be
\label{eq:surf_current}
j_{\mathcal{B}}^i
=
\frac{2}{g_\star^2}\mathcal{E}_{(+)}^i \, .
\ee

Let us note that it is possible to maintain a $\mathbb{Z}_2$ symmetry of the metric without imposing this symmetry on the gauge field. That is because the metric depends only on the square of the charge, while the gauge field depends on its sign as well. Hence, by flipping the sign of the charge across the brane it would in principle be possible to construct configurations that solve the Einstein-Maxwell equations with a brane but for which the surface current vanishes. However, this would then spoil the $\mathbb{Z}_2$ symmetry of the joint configuration. We shall not pursue this further here. 

\subsubsection{The theory on the brane}

Before turning to a discussion of the individual branes, let us describe the three-dimensional theory whose solutions these brane configurations provide. All branes we consider are umbilic, so the discussion applies equally well to each of the subsequent cases. The only difference between cases is how the theory parameters are related to bulk parameters.

Following the standard logic of braneworld holography~\cite{Emparan:2002px, Emparan:2020znc}, we integrate out the bulk on each side of the brane. The on-shell bulk action of either copy, regulated at the brane, equals the effective action $W[h, \mathcal{A}]$ of the holographic CFT living on the brane, minus a series of local terms. In standard holographic renormalization, these local terms would be subtracted so that the CFT effective action is finite when the cutoff is removed. Instead, in braneworld holography, these terms are not subtracted and they become genuine terms in the brane action, where they combine with the tension. Because we keep all such terms, including those terms that would be finite if the cutoff was removed, there is some freedom to split which terms are included in $W$ and which terms are included in the brane effective action (i.e.~a choice of renormalization scheme). We fix all the local terms in the same way, by demanding that they are generated by the local solution of the radial Hamilton-Jacobi equation, order by order in derivatives. This reproduces the standard counterterms of holographic renormalization~\cite{deHaro:2000vlm}. Including the coupling to the brane current, the result is
\begin{align}
    I_\mathcal{B} =&\, \frac{1}{16 \pi G_3} \int_\mathcal{B} \dd^3 x \sqrt{-h} \left[R + \frac{2}{L_3^2} + \ell_4^2 \left(\frac{3}{8} R^2 - R_{ij}R^{ij}\right) + \cdots \right] 
    \nonumber \\
    &- \frac{1}{4 g_3^2} \int_\mathcal{B} \dd^3 x \sqrt{-h} \left[F_{ij}F^{ij} + \cdots  \right] +\int_\mathcal{B} \dd^3 x \sqrt{-h} \mathcal{A}_i j_\mathcal{B}^i + I_{\rm CFT}[h, \mathcal{A}] \, .
\end{align}
Here the ellipsis denote additional higher-derivative and non-minimally coupled terms that are suppressed by further powers of $\ell_4$. The theory parameters of the action are related to the four-dimensional ones according to 
\be 
G_{3} = \frac{G_4}{2 \ell_4} \,, \qquad \frac{1}{L_3^2} = \frac{2}{\ell_4^2} \left(1- \frac{\ell_4}{\ell} \right) \,, \qquad \frac{1}{g_3^2} = \frac{5 \ell_4}{2 g_\star^2} \, .
\ee
Here $I_{\rm CFT} = 2 W$ is the effective action of the cutoff CFT that receives corrections from both sides of the brane. It therefore has central charges $2 c_3$ and $2 c_J$ and its UV cutoff is set by $\ell$. The CFT is coupled to both the brane metric and the restriction of the gauge field to the brane, both of which are dynamical. The bare cosmological length scale $L_3$ differs from the actual AdS$_3$ radius $\ell_3$ on the brane, 
\be 
L_3^2 = \frac{\ell_3^2}{2} \left(1 + \frac{\ell_3}{\sqrt{\ell_3^2 + \ell^2}}\right) \, ,
\ee
where $L_3 \approx \ell_3$ when $\ell/\ell_3 \ll 1$.

\subsubsection{$x = 0$ brane: charged quantum black hole}

The surface $x= 0 $ is always an umbilic hypersurface, since $G(0) = 1$ and $G'(0) = 0$. We retain the region $x \ge 0$ and take the outward-pointing normal on each copy to be
\be 
n_{(x)}^\mu = - A(x-y) \sqrt{G(x)} \, \delta_x^\mu \, .
\ee
With the convention for the extrinsic curvature introduced above, direct evaluation  gives
\be 
K_{ij}^{(x)} = A h_{ij}^{(x)} \, .
\ee
For the discussion of the $x = 0$ brane it is therefore useful to define 
\be 
\ell = \frac{1}{A} \, ,
\ee
after which the tension of the brane takes the form~\eqref{eq:umbilic_tension}. The constraint between solution parameters then gives
\be 
\frac{1}{\ell_4^2} = \frac{1}{\ell^2} + \frac{1}{\ell_3^2}  \qquad \Rightarrow \qquad \ell_3 = \frac{1}{A \sqrt{\lambda}} \, .
\ee
We assume that $\lambda > 0$, ensuring the brane has an asymptotically AdS$_3$ geometry.

The induced metric and gauge potential on the brane are
\begin{align}
    \dd s^2_{\mathcal{B}} &= \frac{1}{A^2 y^2} \left[H(y) \dd t^2 - \frac{\dd y^2}{H(y)} + \dd \phi^2 \right] \,,
    \\
    \mathcal{A} &= \frac{2 e}{A \ell_\star} \, (y-y_-) \, \dd t \, .
\end{align}
The geometry of the brane becomes more transparent upon introducing coordinates,
\be 
y = - \frac{\ell}{r} \, , \quad t = \frac{\tilde{t}}{\ell} \, ,
\ee
in terms of which we arrive at the metric 
\begin{align} 
\dd s^2_{\mathcal{B}} &= - f(r) \dd \tilde{t}^2 + \frac{\dd r^2}{f(r)} + r^2 \dd \phi^2  \, , \qquad f(r) = \frac{r^2}{\ell_3^2} - \kappa - \frac{\mu \ell}{r} + \frac{q^2 \ell^2}{r^2} \, ,
\end{align}
and gauge field
\be 
\mathcal{A} = \frac{2 e \ell}{ \ell_\star} \left(\frac{1}{r_+} - \frac{1}{r}\right) \,  \, \dd \tilde{t}   \quad \Rightarrow \quad F_{\mathcal{B}} = \frac{2 e \ell}{\ell_\star r^2} \dd r \wedge \dd \tilde{t} \, .
\ee
This is the charged qBTZ geometry~\cite{Climent:2024nuj}. 
Here $q^2 = e^2 + g^2$ has retained its definition from earlier, while we have introduce the parameter $\kappa = - k$ to match existing literature on the charged qBTZ solution~\cite{Climent:2024nuj}, and to match the notation for the boundary of the bulk soliton solution in \eqref{solbdy} in the zero backreaction limit.  The coordinate $r$ in range $(0, \infty)$ describes the entire brane geometry; the coordinate $\phi$ is not $2\pi$-periodic in this coordinate system, a fact we shall return to later. 

Let us now consider the Maxwell junction conditions for this brane. Imposing $\mathbb{Z}_2$ symmetry on the gauge field, the electric and magnetic charge parameters must be equal on each side of the brane, {\it i.e.}~ $e_- = e_+$ and $g_- = g_+$.  
The normal flux on either side of the brane is
\be 
\mathcal{E}^i_{(\pm)} = n_\mu^{(\pm)} F^{\mu i}_{(\pm)} = - \frac{2 g \ell}{\ell_\star r^3} \delta_\phi^i \, .
\ee
Hence, the Maxwell junction condition~\eqref{eq:surf_current} gives,
\be  \label{eq:xcurr}
j_{\mathcal{B}}^i =  -\frac{4 g \ell}{g_\star^2 \ell_\star r^3}  \delta_\phi^i \, .
\ee

As in the uncharged case \cite{Hennigar:2026mxv}, the $x=0$ brane intersects the horizon at $y = y_-$, so this corresponds to a compact event horizon ending on the brane with the topology of a disc. If $H(y)$ also has positive roots, the smallest positive root $y=y_+$ is an acceleration horizon, with the topology of a disc with $y_+ \leq x \leq x_+$; $x=y_+$ is a circle on the asymptotic boundary and $x=x_+$ is the axis.

\subsubsection{$y = 0$ brane: charged quantum soliton}

We can also place a brane at the surface $y = 0$, as $H(0) = -\lambda$ and $H'(0) = 0$. We retain the region $y \le 0$. The outward-pointing unit normal to a surface of constant $y$ is 
\be 
n_{(y)}^\mu = A (x-y) \sqrt{-H(y)} \delta_y^\mu \, .
\ee
With our convention for the extrinsic curvature, direct evaluation at $y = 0$ yields
\be 
K_{ij}^{(y)} = A \sqrt{\lambda} h_{ij}^{(y)} \, .
\ee
It is therefore natural for this brane to define 
\be 
\ell = \frac{1}{A \sqrt{\lambda}} 
\ee
so that the tension once again takes the form of eq.~\eqref{eq:umbilic_tension}. With this definition, the relation between bulk and brane scales becomes
\be 
\frac{1}{\ell_4^2} = \frac{1}{\ell^2} + \frac{1}{\ell_3^2} \qquad \Rightarrow \qquad \ell_3 = \frac{1}{A} \, .
\ee

The induced metric and gauge potential on the brane are
\begin{align} \label{qsolx}
\dd s^2_{\mathcal{B}} &= \frac{1}{A^2 x^2} \left[ - \lambda \dd t^2 + \frac{\dd x^2}{G(x)} + G(x) \dd \phi^2 \right]\,,
\\
\mathcal{A} &= \frac{2 g}{A \ell_\star} \, (x-x_+)\, \dd \phi \, .
\end{align} 
Defining coordinates and parameters,
\be 
x = \frac{\ell_3}{r} \, , \quad \phi = \frac{\varphi}{\ell_3} \, , \quad t = \frac{\tau}{\sqrt{\lambda}} \,, \quad \kappa = k \, , \quad  \hat{\mu} = \sqrt{\lambda} \mu \,, \quad \hat{q}^2 = \lambda q^2 \, ,
\ee
then the induced geometry takes the form of the charged quantum soliton,
\begin{align} \label{qsol}
    \dd s_{\mathcal{B}}^2 &= - r^2 \dd \tau^2 + \frac{\dd r^2}{f(r)} + f(r) \, \dd \varphi^2 \, ,
    \\
    \mathcal{A} &=  \frac{2 g \ell_3}{\ell_\star} \left(\frac{1}{r} - \frac{1}{r_0}\right) \dd \varphi \, . 
\end{align}
where 
\be 
f(r) = \frac{r^2}{\ell_3^2} - \kappa - \frac{\ell \hat{\mu}}{r} - \frac{\ell^2 \hat{q}^2}{r^2} \, .
\ee 
Note that the relation between $\kappa$ and $k$ here is different from the $x=0$ brane. These conventions match the ones used for the boundary of the AdS black hole in \eqref{eq:zero-backreaction-qsol}. This reflects we don't change $\kappa$ in the double analytic from the black hole to the soliton, while the double analytic continuation of the C-metric maps it to itself with $k \to -k$. At a point $r_0$ where the metric function $f(r_0) = 0$ vanishes, $\varphi$ must be periodically identified to ensure regularity. On the brane, $r \in [r_0, \infty)$ covers the full brane geometry (but not the bulk). 

Imposing $\mathbb{Z}_2$ symmetry, we take the charges to be identical on both copies. Hence, the tangential components of the field strength automatically agree. The normal flux is seen to be
\be 
\mathcal{E}_{(\pm)}^i = - \frac{2 e \ell_3}{\ell_\star \, r^3} \delta_\tau^i \, .
\ee
Satisfying the Maxwell junction condition therefore requires the presence of a surface current,
\be  \label{eq:ycurr}
j_{\mathcal{B}}^i = - \frac{4 e \ell_3 }{g_\star^2 \ell_\star \, r^3} \delta_\tau^i \, .
\ee

The $y=0$ brane does not intersect horizons in the bulk, as the horizons also lie at constant values of $y$. In the presence of the $y=0$ brane, we retain the negative $y$ region, so any acceleration horizons are irrelevant. We have a horizon at $y=y_-$, which is either a non-compact horizon with a boundary on the asymptotic boundary or a compact black hole horizon depending on the root structure.

\subsubsection{Configurations with two branes}

As discussed in \cite{Hennigar:2026mxv}, for a single brane setup in the uncharged case, the brane can be brought to $x=0$ (or $y=0$) by a shifting and scaling of the coordinates and parameters, but there exist inequivalent two-brane setups which are not related by a shifting and scaling of the coordinates and parameters. For example, while inserting branes  at $x=0$ and $y=0$ is physically equivalent to inserting branes at $x=\xi_c$ and $y=\xi_c$ (where $\xi_c$ is a stationary point of $G(x)$ or $H(y)$), it is not equivalent to inserting branes at $x=0$ and $y=\xi_c$. This is because $x$ and $y$ must shift and scale by the same amount when mapping between different two-brane setups. 

The situation in the charged case is similar, but there are more stationary points to consider; in the uncharged case $G$ and $H$ had two stationary points, while the charged case they have three (in all the physically relevant cases where $H$ has negative roots). If we have a single brane, a coordinate transformation and a parameter rescaling can bring the brane to $x=0$ (or $y=0$), but when we consider inserting two branes, we can put them at different stationary points. There are then nine possible two-brane configurations, but some of them are physically equivalent to each other. In particular, any configuration where the $x$ and $y$ branes are inserted at the same stationary point is physically equivalent to the $x=0$ and $y=0$ brane configuration, since $x$ and $y$ can be shifted and rescaled to bring the branes to this position. 

For configurations with a brane at $x=\xi_1$ and $y=\xi_2$, we retain the spacetime region with $x > \xi_1$ and $y < \xi_2$. For $\xi_1 \neq \xi_2$, the most significant physical distinction is between cases with $\xi_1 < \xi_2$, where we retain a portion of the conformal boundary at $x=y$, and cases with $\xi_1 > \xi_2$, where the branes intersect in the bulk. As in \cite{Hennigar:2026mxv}, we will not explore these cases with branes at different stationary points in detail, focusing on the case with two branes at $x=0$ and $y=0$. With these two branes, the bulk spacetime has a single compact event horizon at $y=y_-$ in all cases, with the topology of a disc ending on the $x=0$ brane. The conformal boundary of the C-metric is almost entirely removed, only the ring at $x = y = 0$ remains.

\subsection{Zero backreaction limit}
From the point of view of the effective theory on the brane, the higher-curvature corrections are controlled by the UV scale $\ell$; the dimensionless parameter is the ratio $\ell/\ell_3$. For the $x=0$ brane $\ell/\ell_3 = \sqrt{\lambda}$, so the zero backreaction limit is $\lambda \to 0$, and for the $y=0$ brane $\ell/\ell_3 = 1/\sqrt{\lambda}$, so the zero backreaction limit is $\lambda \to \infty$. From the bulk point of view, these are limits in which the charged C-metric simplifies, and the relevant brane goes off to the boundary. For $\lambda \to \infty$ the charged C-metric \eqref{ch_cmetric_xy} reduces to the charged black hole \eqref{rnAdS}, and the $y=0$ brane goes off to the boundary, while for $\lambda \to 0$ the charged C-metric reduces to the soliton \eqref{solads}, and the $x=0$ brane goes off to the boundary. 

These limits are most easily understood by rescaling the $y,t$ coordinates and taking a factor of $\lambda$ out of $H(y)$, defining 
\begin{equation}
    y = \sqrt{\lambda} \bar y, \quad t = \frac{\bar t}{\sqrt{\lambda}}, \quad H(y) = - \lambda \bar F(\bar y). 
\end{equation}
With these definitions, and recalling $A^{-2} = \ell_4^2 (\lambda +1)$, the metric \eqref{ch_cmetric_xy} becomes 
\begin{equation}
\mathrm{d}s^2
=
\frac{\ell_4^2 (\lambda +1) }{(x- \sqrt{\lambda} \bar y)^2}
\left[ -
\bar F(\bar y)\mathrm{d}\bar t^2
+\frac{\mathrm{d} \bar y^2}{\bar F(\bar y)}
+\frac{\mathrm{d}x^2}{G(x)}
+G(x)\mathrm{d}\phi^2
\right],
\end{equation}
with gauge potential
\begin{equation}
\mathcal A
=
\frac{2 \ell_4 \sqrt{\lambda +1}}{\ell_\star}
\left(
e (\bar y- \bar y_-)\mathrm{d} \bar t
+g (x-x_+) \mathrm{d}\phi
\right) ,
\end{equation}
and 
\begin{equation}
\label{cmetric_structure_functions}
\bar F(\bar y)
=
1+k \bar y^2+ \mu \sqrt{\lambda} \bar y^3+ q^2 \lambda \bar y^4. 
\end{equation}

Taking $\lambda \to 0$ at fixed $\mu, e, g$, the $\bar t, \bar y$ part of the metric becomes independent of the parameters $\mu, e,g$, and the overall conformal factor becomes $\ell_4^2/x^2$. Setting $x = \frac{\ell_4}{\rho}$, $\phi = \Phi$ and $G(x) = x^2 \mathcal F(\rho)$, this becomes the soliton metric \eqref{solads}. Infinity in the C-metric is at $x = \sqrt{\lambda} \bar y$ in these  coordinates, so points at $x=0$ for fixed $\bar y$ go to infinity as we take the limit. The surface current on the $x=0$ brane \eqref{eq:xcurr} is in these coordinates
\begin{equation}
    j_{\mathcal{B}}^i =  \frac{4 g \lambda^{3/2} \bar y^3}{g_\star^2 \ell_\star \ell_4^2 (\lambda+1)}  \delta_\phi^i \, ,
\end{equation}
so this vanishes as $\lambda \to 0$. 

In the limit as $\lambda \to \infty$, we should hold $\hat \mu = \sqrt{\lambda} \mu$, $\hat e = \sqrt{\lambda} e$, $\hat g = \sqrt{\lambda} g$ fixed so that $\bar F(\bar y)$ and $\mathcal A$ have finite limits. Then the $x,\phi$ part of the metric becomes independent of $\hat \mu, \hat e, \hat g$ and the overall conformal factor becomes $\ell_4^2/\bar y^2$. Setting $\bar{y} = -\frac{\ell_4}{\rho}$, $\bar t = \tau$, and $\bar F(\bar y) = \bar y^2 \mathcal F(\rho)$, this becomes the charged black hole metric \eqref{rnAdS}. Points at $\bar y=0$ for fixed $x$ go to infinity as we take the limit. The surface current on the $y=0$ brane \eqref{eq:ycurr} is in these coordinates
\begin{equation}
 j_{\mathcal{B}}^i = - \frac{4 \hat e x^3}{g_\star^2 \ell_\star \ell_4^2  \sqrt{\lambda} (\lambda+1) \, } \delta_\tau^i \, , 
\end{equation}
so this vanishes as $\lambda \to \infty$. 
\FloatBarrier 
\section{Quantum soliton}
\label{sec:qsol}

Physically, we understand the $y=0$ brane geometries \eqref{qsol} in the C-metric as backreacted versions of the boundary geometry \eqref{eq:zero-backreaction-qsol}. We refer to this backreacted geometry as the quantum soliton.  The backreaction describes the effects of the quantum matter and backgrounds (\ref{adsbdyTJ},\ref{eq:ads_A0}) when we introduce a finite gravitational coupling on the branes, by moving from the $\lambda \to \infty$ limit to finite $\lambda$.  In the charged case, this involves both the backreaction of the CFT state on the geometry and that of the background magnetic field. The result depends on the strength of the backreaction, encoded in $\lambda$, and the CFT state and field, determined by the parameters $\hat \mu$, $\hat q$. The metric on the brane is determined by the function $G(x)$, as seen in the original form \eqref{qsolx} of the geometry on the brane. From this we see that the backreaction depends on the CFT parameters and the strength of the backreaction only through the combinations
 $\mu = \frac{\hat \mu}{\sqrt{\lambda}}$, $q = \frac{\hat q}{\sqrt{\lambda}}$, the parameters appearing in the bulk C-metric solution. 

 One might expect that the effects of backreaction on the brane solution will be a complicated function of the parameters $\mu, q$. However, in the regime of large but finite $\lambda$ where the effective theory on the brane is reliable, there is a significant simplification, because of the constraint $\hat \mu \geq \hat \mu_{\rm ext}$ for the existence of a horizon in the bulk.  We are interested in the regime of large $\hat \mu$, $\hat q$, to compensate for the weakness of the backreaction from large $\lambda$. In this regime the extremal bound \eqref{kpbound} simplifies to \eqref{largeqconst}. Thus in this regime the sub-extremality constraint is non-linear, and in terms of $\mu, q$, it becomes  
\begin{equation}
    q <  \frac{\sqrt{3}}{2^{4/3}} \frac{\mu^{2/3}}{\lambda^{1/6}}. 
\end{equation}
Thus, for large values of $\lambda$, $q$ is parametrically suppressed compared to $\mu$ for there to be a horizon in the bulk. This implies that the backreaction on the geometry  of the brane is controlled by $\mu$, and is the same as in the uncharged case discussed in \cite{Hennigar:2026mxv}. 

As in the uncharged case \cite{Hennigar:2026mxv}, the quantum soliton solutions we have obtained describe two scenarios: for $\kappa=1$ the quantum soliton describes the backreaction of a charged thermal gas on global AdS$_3$, while for $\kappa=-1$ it describes the backreaction on a geometry with a horizon, where the state of the matter is not regular on the horizon. 

\subsection{Backreacted global AdS$_3$}

In the former case, the bulk solution in the zero backreaction limit is a Reissner-Nordstr\"om AdS$_4$ black hole \eqref{rnAdS} with a spherical horizon. In the zero backreaction limit, the $y=0$ brane becomes half of the conformal boundary, described in global AdS$_3$ coordinates. The bulk black hole gives a charged thermal CFT state on the boundary. 

In this $\kappa=1$ case, the function $f(r)$ had a smooth origin before backreaction, see \eqref{eq:zero-backreaction-qsol}. The effect of the backreaction is to shift the location of the origin and to change the period $\Delta \varphi$. The latter alters the ADM mass of the solution as read off at infinity on the brane.  Writing $\Delta \varphi = 2\pi \Delta$, the mass of the soliton is $M_3 = -\Delta^2$, in units where global AdS$_3$ has $M_3=-1$. As noted above, the effect of backreaction is controlled by the mass parameter $\mu$, and $M_3$ is a monotonically increasing function of $\mu$, with $M_3 \to -1$ as $\mu \to 0$ and $M_3 \to 0$ as $\mu \to \infty$.  

\subsection{The disappearing horizon}

For $\kappa=-1$, the bulk solution in  the zero backreaction limit is a Reissner-Nordstr\"om AdS$_4$ black hole with a hyperbolic horizon. The relation to the brane in the C-metric is established by writing the boundary metric in the Rindler-AdS$_3$ form \eqref{eq:zero-backreaction-qsol}, which for $\kappa=-1$  has a Rindler-like horizon at $r=0$ with temperature $T_{\rm R} = 1/(2\pi)$.  In this case the CFT stress tensor, current and background magnetic field are not regular on the horizon at $r=0$ unless $\mu=e=g=0$. The solution at finite $\lambda$ thus gives us insight into how such a non-regular state backreacts on the geometry. 

At finite $\lambda$, we obtain a backreacted solution in which the horizon at $r=0$ is replaced by a smooth origin at $r_0$, where $f(r_0)=0$. The CFT state is regular everywhere on the backreacted solution. In \cite{Hennigar:2026mxv}, we discussed the uncharged case, where $f(r)$ has a positive root for $\mu >0$ (corresponding to states hotter than the Rindler horizon, $T_{\rm H} > T_{\rm R}$), but not for $\mu <0$. Here we see that for non-zero charge $q$, the backreacted solution will have a positive root for $f(r)$ for all values of $\mu$. This is a significant difference in the current charged case. It means that we have a model of the backreaction of the CFT state for the full parameter space of hyperbolic black holes shown in Figure~\ref{fig:rn-params}, including states colder than the Rindler horizon. As discussed in Section~\ref{sec:ads-rindler}, these colder states fall into two categories: states of negative mass with $\hat{\mu} < 0$, and a narrow band of states with positive mass with $0 \le \hat{\mu} \le \hat{\mu}_{\rm R}.$ These two regimes behave very differently with respect to the three-dimensional effective theory on the brane.

Consider first the solitons with negative mass $\hat{\mu} < 0$, where every state satisfies $T_{\rm H} < T_{\rm R}$. From the extremality bound~\eqref{kpbound}, negative values  $\hat \mu < 0$ exist only for $\hat q < \frac{1}{2}$, and possible negative values of $\hat \mu$ are bounded below, $\hat \mu \geq - \frac{2}{3 \sqrt{3}}$. Thus, we can have $\hat \mu <0$ only for $\mathcal O(1)$ values of $\hat \mu$ and $\hat q$. This implies that the smooth origin $r=r_0$ such that $f(r_0)=0$ is at small $r$, 
\begin{equation}
    \frac{r_0}{\ell} \approx  \frac{-|\hat \mu| + \sqrt{\hat \mu^2 + 4 \hat q^2}}{2}.
\end{equation}
Since $r_0 \sim \ell$, the UV cutoff scale of the effective theory on the brane, this resolution of the singularity is outside of the regime of validity of the effective brane theory.\footnote{This is analogous to the branch of quantum BTZ black holes that limit to conical singularities when the backreaction vanishes. In that case, the horizon size near the zero-backreaction limit is proportional to $\ell$~\cite{Emparan:2020znc}.} The resulting ADM mass on the brane is also small in this $\hat \mu <0$ regime, 
\begin{equation}
    M_3 \approx \frac{(-|\hat \mu| + \sqrt{\hat \mu^2 + 4 \hat q^2})^4} {4\lambda (\hat \mu^2 + 4 \hat q^2)}. 
\end{equation}
Thus $M_3 \sim \tfrac{1}{\lambda}$, starting from zero at $\hat q=0$ and reaching $M_3 = \frac{1}{4\lambda}$ at $\hat q = \frac{1}{2}$, where $\hat \mu = 0$.

The situation is qualitatively better in the positive mass regime, because the colder band is not restricted to small values of the parameters. For the backreaction to be visible on scales at which the brane theory is still reliable, we need $\mu = \hat{\mu}/\sqrt{\lambda}$ to be ${\cal O}(1)$, that is $\hat{\mu} \sim \sqrt{\lambda}$. It turns out for positive masses the band of colder states reaches such values of $\hat{\mu}$. As shown in section~\ref{sec:ads-rindler}, the dashed curve $T_{\rm H} = T_{\rm R}$ is parametrized by $u_0 \ge 1$ as 
\be 
\hat{\mu}_{\rm R} = 4 u_0^3 - 2 u_0^2 - 2 u_0 \, ,\qquad \hat{q}^2 = 3 u_0^4 - 2 u_0^3 - u_0^2 \, ,
\ee
which extends to arbitrarily large $u_0$. For $u_0 \gg 1$ we have $\hat{\mu}_{\rm R} \approx 4 u_0^3$ and $\hat{q} \approx \sqrt{3} u_0^2$. Choosing $u_0 \sim \lambda^{1/6}$ then gives the required scaling $\hat{\mu} \sim \sqrt{\lambda}$ with $\hat{q} \sim \lambda^{1/3}$ so that $q \sim \lambda^{-1/6}$. The smooth origin then sits at $r_0 \sim \ell_3 \gg \ell$, well within the regime of validity of the three-dimensional effective theory. 

Because $q$ is suppressed relative to $\mu$, these colder states backreact on the geometry in the same way as the uncharged states of~\cite{Hennigar:2026mxv}. To see this explicitly, we can write the metric function $f(r)$ in terms of the C-metric parameters
\be 
f(r) = \frac{r^2}{\ell_3^2} + 1 - \frac{\mu \ell_3}{r} - \frac{q^2 \ell_3^2}{r^2} \, .
\ee
Near the origin, $r \sim \ell_3$, and the charge-dependent term is suppressed by $q^2 \sim \lambda^{-1/3}$ relative to the other terms. This means that the the leading order behaviour of $f(r)$ is the same as the uncharged soliton with the same value of $\mu$. This leads to a rather striking behaviour. Consider two states with the same ${\cal O}(1)$ value of $\mu$: a near-extremal charged state from the colder band of states with $T_{\rm H} < T_{\rm R}$, and an uncharged state with $T_{\rm H} \approx 3 \hat{\mu}^{1/3}/(4 \pi) \sim \lambda^{1/6} \gg T_{\rm R}$. Despite the fact that the temperatures of these two states differ by a large factor, they produce the same brane geometry to leading order. This is possible because the backreaction of the state is determined by its stress tensor, which is controlled by $\mu$, rather than by the temperature. 
\section{Euclidean action and thermodynamics}
\label{sec:thermo}

As in the uncharged case discussed in \cite{Hennigar:2026mxv}, the calculation of the Euclidean action and the derivation of a first law of thermodynamics for the black hole is most easily analysed in a two-brane scenario, with branes at $x=0$ and $y=0$. This is because variations in the parameters of the C-metric induce variations in the geometry of the conformal boundary, so there will generically be additional work terms in the first law associated with deformations of the boundary geometry~\cite{ThermoForthcoming}. The two-brane scenario avoids this complication because the introduction of both branes removes the conformal boundary from the spacetime, apart from the circle at $x=y=0$. 

We carry out the calculation of the Euclidean action in this case in appendix \ref{action}. The result is \eqref{eq:total-src}, 
\begin{equation} \label{eq:action}
 I_{E}
 =\frac{\Delta\tau\Delta\phi}{8\pi G_4A^2}
 \left[\frac{1+\lambda}{(x_+-y_-)^2}
 -\frac1{x_+^2}-\frac{\lambda}{y_-^2}
 -2(e^2-g^2)x_+y_-\right].
\end{equation}
The first part is the gravitational action, which has the same form as in the uncharged case. The final term is the contribution from the Maxwell action and the current terms on the branes. 

\begin{figure}
    \includegraphics[width=0.5\linewidth]{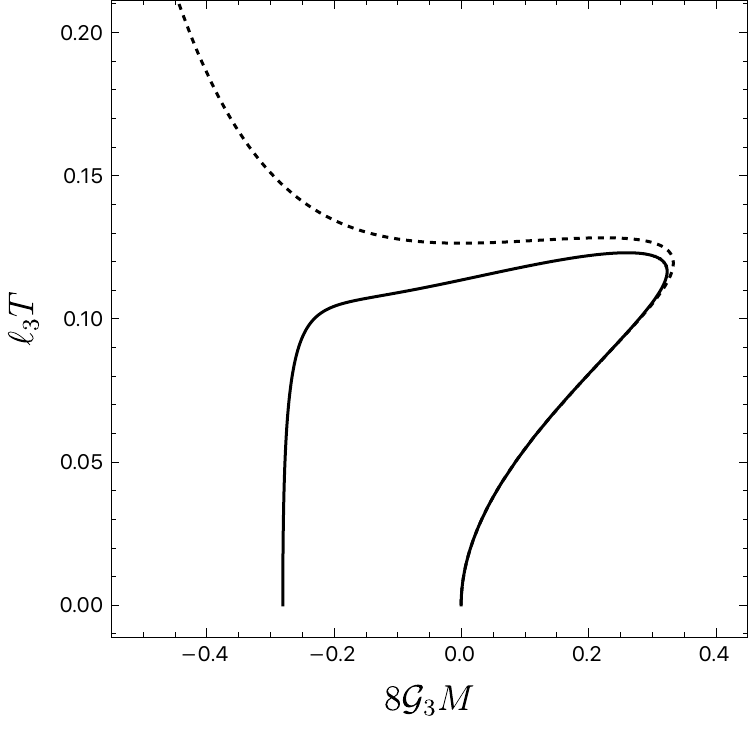}
    \quad \includegraphics[width=0.51\linewidth]{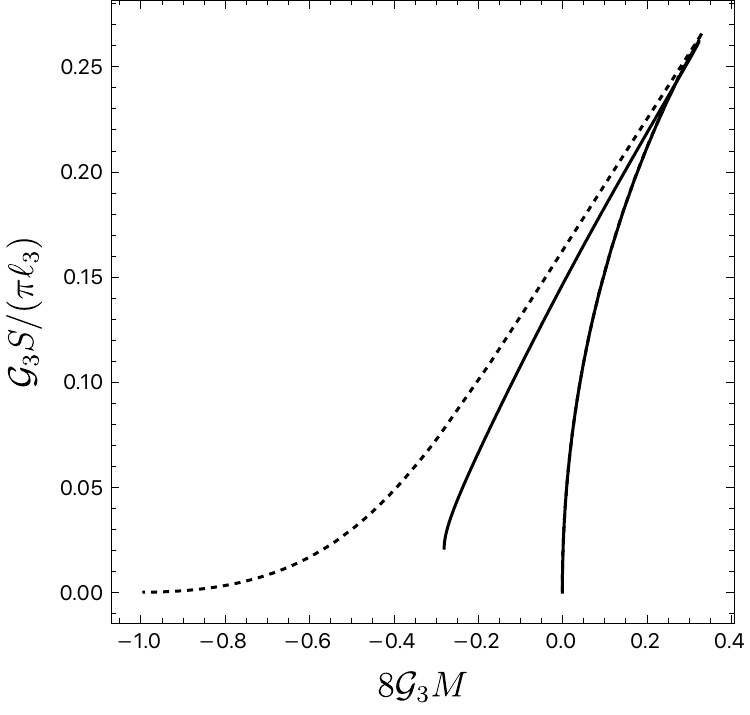}
    \caption{Temperature (left) and entropy (right) of the soliton as a function of the mass for $\lambda=\frac{1}{4}$. The solid lines describe the charged quantum soliton with $\gamma=0.3$, while dashed lines describe neutral solitons with $\gamma=0$ (the dashed line lies on top of the solid line on the right). The solutions with $M<0$ correspond to the $\kappa=1$ branch, while $M>0$ corresponds to the $\kappa=-1$ branch. The charged bulk black holes become extremal at a certain $M<0$, where the temperature becomes zero, but the horizon area is finite. This accounts for the finite entropy (right) of the charged soliton at zero temperature (left). }
    \label{fig:TS}
\end{figure}

The first law for the $x=0$ brane was described in \cite{Feng:2024uia,Climent:2024nuj}, so we focus here on describing the first law from the perspective of the $y=0$ brane. We write  
\begin{equation}
    \Delta= \frac{\Delta\phi}{2\pi}=\frac{2}{|G'(x_+)|}.
\end{equation}
In order to compute the mass and the temperature, we normalise the angular period to $2\pi$ and the time to global AdS$_3$ time at infinity, by setting
\begin{equation}
    r=\frac{\Delta}{A x},\qquad t_s=\frac{\sqrt{\lambda}}{A\Delta}t,\qquad \tilde\phi=\frac{\phi}{\Delta},\qquad \tilde\phi\sim \tilde\phi+2\pi,
\end{equation}
which brings the metric to the form 
\begin{equation}
    \dd s^2=-A^2 r^2 \dd t_s^2+\frac{\dd r^2}{f(r)}+\frac{f(r)}{A^2}\dd \tilde\phi^2
\end{equation}
where \begin{equation}
    f(r)=A^2 r^2-\kappa \Delta^2-\frac{\mu\Delta^3}{A r}-\frac{q^2\Delta^4}{A^2 r^2} , 
\end{equation}
with $\kappa = k$. The regularity of the bulk horizon fixes the periodicity of Euclidean time $\tau_s=i t_s$, $\tau_s \sim \tau_s + \beta_{\rm sol}$, which gives rise to the soliton temperature $T_{\rm sol} = \beta_{\rm sol}^{-1}$.
It will be useful to introduce some auxiliary variables
\begin{equation}
    \nu=\frac{1}{\sqrt{\lambda}}=\frac{\ell}{\ell_3},\qquad \qquad z=-\frac{\nu y_{-}}{x_{+}},
\end{equation}
\begin{equation}
    \gamma_{e}=e x_{+}^2,\qquad \gamma_g=g x_{+}^2,\qquad \gamma^2=\gamma_e^2+\gamma_g^2.
\end{equation}
The temperature, ADM mass and the entropy are then
\begin{eqnarray}
    T_{\rm sol}&=&\frac{z \nu^2(2\nu+3z+z^3)-\gamma^2 z^4(\nu+z)^2}{2\pi \ell_3 \nu [\nu (\nu+3\nu z^2+2z^3)+\gamma^2 z^2(\nu+z)^2]},\\
    M_{\rm sol} &=& -\frac{\kappa \Delta^2}{8\mathcal{G}_3}=\frac{\nu z^2(\nu+z)[\nu(\nu-z^3)+\gamma^2 z^3(\nu+z)]}{2\mathcal{G}_3 [\nu(\nu+3\nu z^2+2z^3)+\gamma^2 z^2(\nu+z)^2]^2},\\
    S_{\rm sol} &=& \frac{\Delta \phi x_+}{2 G_4 A^2 (-y_-)(x_+-y_-)}=\frac{\pi \ell_3 \nu^2 z}{\mathcal{G}_3 [\nu(\nu+3\nu z^2+2z^3)+\gamma^2 z^2(\nu+z)^2]}. 
\end{eqnarray}
where $\mathcal{G}_3$ is the renormalised Newton constant given by, $\mathcal{G}_3=\frac{\ell_4}{\ell}G_3=\frac{G_4}{2\ell}$. In figure \ref{fig:TS}, we plot the temperature and entropy as a function of mass for the charged soliton, comparing to the behaviour of the uncharged soliton. Note that the plot is for an intermediate value of $\lambda$, where the effects of charge are not as simple as in our discussion in the previous section.

\begin{figure}
    \centering
    \includegraphics[width=0.5\linewidth]{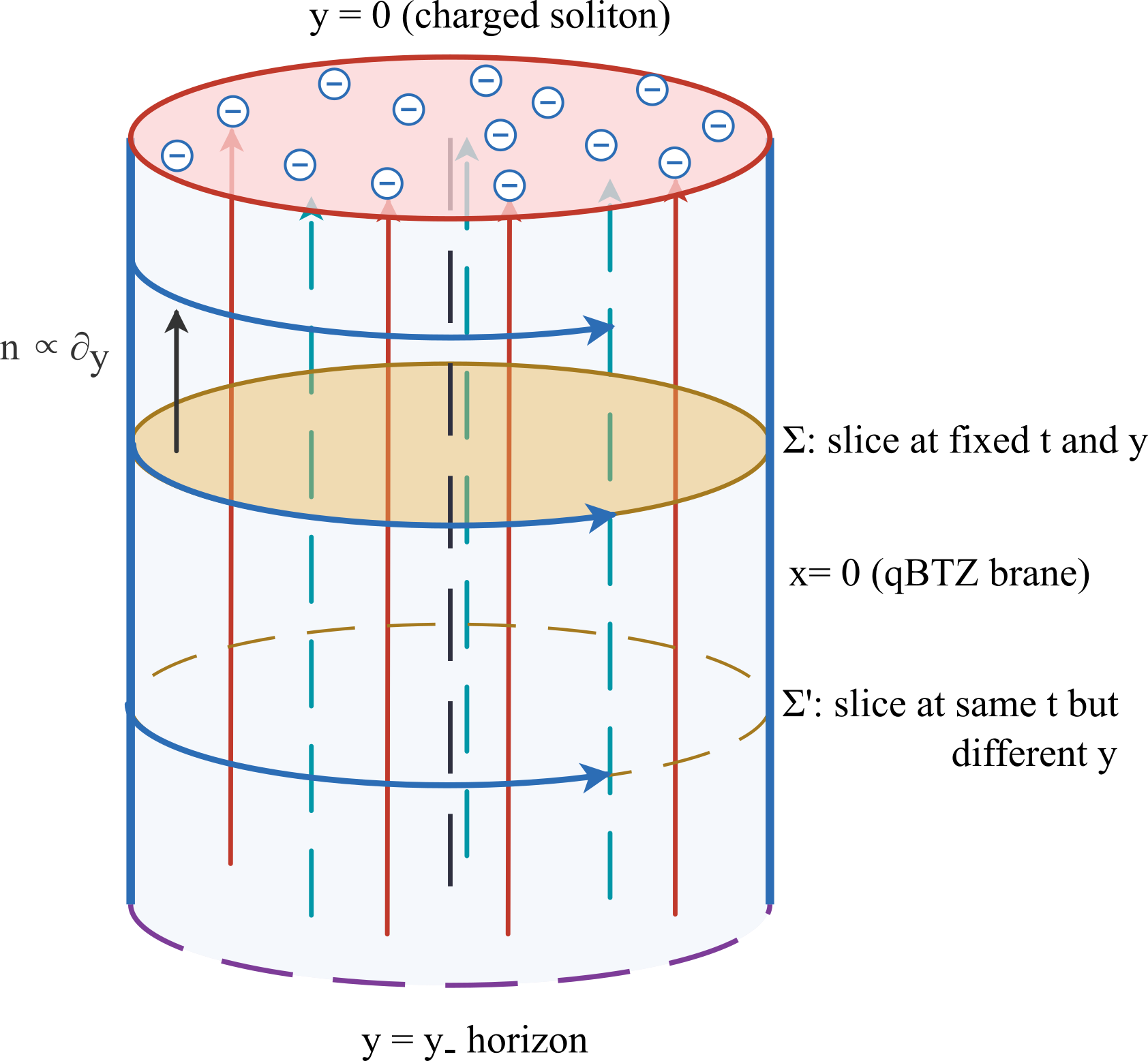}
    \caption{A constant time slice of the bulk with two branes inserted at $x=0$ (blue surface) and $y=0$ (red disk) in $(x,y,\phi)$ coordinates for $e,g>0$. The solid red lines are the electric field lines and the dashed cyan lines are the magnetic field lines. The electric field lines end on the soliton brane at $y=0$, and the magnetic field is sourced by the surface current on $x=0$ brane (blue arrows). The unit vector $n\propto \partial_y$ is normal to the disk $\Sigma$ and points towards increasing $y$. The black dashed line is the rotational axis $x=x_+$. The charges $Q_e$ and $Q_g$ are the fluxes of $\star F$ and $F$ through $\Sigma$, and they are the same through $\Sigma'$. Although we have shown one copy of the bulk, our full setup is $\mathbb{Z}_2$ symmetric, where two of these copies are glued across the two branes.}
    \label{fig:sigma}
\end{figure}

These expressions for the mass and entropy are obtained from the fall-off along the brane and the area of the black hole horizon in the bulk. They can also be obtained from the Euclidean action \eqref{eq:action}. We have Neumann boundary conditions for the gauge field on the branes, fixing the normal derivative of the gauge field, which is related to the surface current in the action \eqref{eq:branecurrent}. The electric and magnetic charges are calculated by integrating $\star F$ and $F$ respectively over a surface $\Sigma$ as shown in figure \ref{fig:sigma}. Thus the boundary conditions on the $y=0$ brane fix the electric charge $Q_e$, and the chemical potential $\mu_g$ dual to the magnetic charge. The Euclidean action thus corresponds to a mixed ensemble, $Z(\beta,\mu_g,Q_e) \approx e^{-I_E}$. Thus 
\begin{equation}
    M=\partial_\beta I_{E}+\mu_g Q_g,\quad\quad S=(\beta \partial_\beta -1)I_E.
\end{equation}

The electric and magnetic charges obtained by integrating over $\Sigma$ are 
\begin{eqnarray}
    Q_e &=&\frac{2}{g_{\star}^2}\int_{\Sigma}\star F= \frac{4 e x_+\Delta \phi}{A g_\star^2\ell_\star}=\frac{\ell_\star \gamma_e z^2 (\nu+z)}{2\mathcal{G}_3(\nu(\nu+3\nu z^2+2z^3)+\gamma^2 z^2(\nu+z)^2)},\\
   Q_g&=&\frac{2}{g_{\star}^2}\int_{\Sigma} F=\frac{4g x_+\Delta \phi}{A g^2_\star\ell_\star}=\frac{\ell_\star \gamma_g z^2(\nu+z)}{2 \mathcal{G}_3 (\nu(\nu+3\nu z^2+2z^3)+\gamma^2 z^2(\nu+z)^2)}, 
\end{eqnarray}
and their conjugate potentials are
\begin{eqnarray}
  \mu_e&=&\xi^{\mu}\mathcal{A}_{\mu}|_{y=0}-\xi^{\mu}\mathcal{A}_{\mu}|_{y=y_-}\\
  \mu_g&=&\xi^{\mu}\tilde{\mathcal{A}}_{\mu}|_{y=0}-\xi^{\mu}\tilde{\mathcal{A}_\mu}|_{y=y_-}
\end{eqnarray}
where $\mathcal{A}_\mu$ is the gauge potential, $\tilde{\mathcal{A}}_\mu$ is the dual potential defined as $d\tilde{\mathcal{A}}=-\star F$ and $\xi^\mu$ is the Killing vector $\xi=\partial_{t_s}=\frac{A\Delta}{\sqrt{\lambda}}\partial_t$.
Evaluating those expressions yields,
\begin{eqnarray}
     \mu_e&=&-\frac{e y_-\Delta \phi}{\pi \ell_\star\sqrt{\lambda}}=\frac{4\nu \gamma_e z^3(\nu+z)}{\ell_\star(\nu(\nu+3\nu z^2+2z^3)+\gamma^2 z^2(\nu+z)^2)},\\
    \mu_g&=&-\frac{g y_- \Delta \phi}{\pi \ell_\star\sqrt{\lambda}}=\frac{4\nu \gamma_g z^3(\nu+z)}{\ell_\star(\nu(\nu+3\nu z^2+2z^3)+\gamma^2 z^2(\nu+z)^2)}.
\end{eqnarray}
These satisfy the first law 
\begin{equation}
    \dd M=T \dd S+\mu_e \dd Q_e+\mu_g \dd Q_g,
\end{equation}
which is obtained by varying the $(z,\gamma_e,\gamma_g)$  keeping $(\nu, \ell_3, \mathcal G_3,\ell_\star)$ fixed.

\section{Discussion}

The AdS C-metric provides a basis for constructing exact, non-perturbative solutions to semiclassical theories of gravity and matter through braneworld holography. Here we used this formalism to obtain charged generalizations of our recently constructed quantum solitons~\cite{Hennigar:2026mxv}.  The technical ingredient that made this possible is the fact that the charged C-metric maps into the same local class of metrics under double analytic continuation, with $k \to -k$, $\lambda \to 1/\lambda$ and an exchange of electric and magnetic charges. This exchanges the charged quantum BTZ brane at $x=0$~\cite{Feng:2024uia,Climent:2024nuj} with the soliton brane at $y=0$.

In the zero backreaction limit, the bulk solutions reduce to the dyonic Reissner--Nordstr{\"o}m--AdS family of metrics, and the brane carries a thermal CFT state at chemical potential $\mu_e$ in a background magnetic field, on global AdS$_3$ or on AdS--Rindler. The charged quantum soliton describes the backreaction of this state on the metric and on the gauge field, which become dynamical on the brane. The most significant qualitative difference arises for AdS--Rindler. Without charge, only states hotter than the Rindler horizon have backreacted counterparts. With charge, every hyperbolic black hole state does, including those colder than the horizon, and in every case the horizon is replaced by a smooth origin. 
\section*{Acknowledgements}

YD is supported by STFC. The work of RAH was supported by a Willmore Fellowship at Durham University and by an STFC Ernest Rutherford Fellowship UKRI5636.  AKP is supported by the European Union’s Horizon Europe research and innovation programme under the Marie Sklodowska-Curie grant agreement No. 101210745. The work of SFR is supported in part by STFC through grant numbers ST/X000591/1 and UKRI4226.

\appendix

\section{Supersymmetry}
\label{sec:susy}

The supersymmetry in the C metric is discussed in detail in \cite{Klemm:2013eca}. They work in a slightly different coordinate system, where the metric is written as 
\begin{equation} \label{KNCmetric}
ds^2 = \frac{1}{(p+q)^2}\left[-Q(q)\,d\tau^2 + P(p)\,d\sigma^2
      + \frac{dp^2}{P(p)} + \frac{dq^2}{Q(q)}\right],
\end{equation}
with 
\begin{equation}
    P(p) = \left(\alpha - \frac{\Lambda}{6}\right) + 2np - \varepsilon p^2
      + 2mp^3 - (Q^2+P^2)p^4, 
\end{equation}
\begin{equation}
    Q(q) = \left(-\alpha - \frac{\Lambda}{6}\right) + 2nq + \varepsilon q^2
      + 2mq^3 + (Q^2+P^2)q^4.
\end{equation}
They find that the integrability conditions which provide necessary conditions for supersymmetry to be preserved are
\begin{equation} \label{KNsusy1}
    m[m^2 - (Q^2+P^2)\varepsilon]+2n(Q^2+P^2)^2=0,   
\end{equation}
\begin{equation} \label{KNsusy2}
    m^2 [ \frac{(Q^2-P^2)}{2\ell_4^2}+(Q^2+P^2)\alpha] + n^2 (Q^2+P^2)^2=0.
\end{equation}

We can set $n=0$ by a translation in the $p, q$ coordinates. With $n=0$, these coordinates are related to our coordinates by
\begin{equation}
    p = Ax, \quad q=  -Ay, \quad \sigma = \frac{\phi}{A}, \quad \tau = \frac{t}{A}, 
\end{equation}
with the parameters related as 
\begin{equation}
    \alpha = A^2 + \frac{\Lambda}{6} = A^2 - \frac{1}{2 \ell_4^2},  \quad \varepsilon = k, \quad m = - \frac{\mu}{2A}, \quad Q = \frac{e}{A}, \quad P = \frac{g}{A}, 
\end{equation}
noting that \cite{Klemm:2013eca} choose units such that $\ell_\star = 2$. 
In terms of our parameters the supersymmetry conditions (\ref{KNsusy1},\ref{KNsusy2}) become (recalling $\frac{1}{A^2 \ell_4^2} = \lambda+1$)
\begin{equation} \label{susy1}
    \frac{\mu}{4A^3}  \left[ \mu^2  - 4(e^2+g^2)k \right]= 0, 
\end{equation}
\begin{equation} \label{susy2}
    \frac{\mu^2}{8A^2}  \left[ e^2 - \lambda g^2 \right] =0. 
\end{equation}
This choice of coordinates thus nicely simplifies the supersymmetry conditions. These conditions could both be satisfied by setting $\mu=0$, or we can set the terms in square brackets to zero. Klemm and Nozawa \cite{Klemm:2013eca} analysed the second case in detail, and showed that the solution with this choice of parameters could be rewritten in the canonical form of \cite{Caldarelli:2003pb}, thus demonstrating the presence of a supersymmetry. 

Setting $\mu=0$ satisfies the integrability conditions, but it is not in general sufficient to ensure supersymmetry of the solution. This is most easily seen by observing that with $\mu=0$, $H(y)  = -\lambda -k y^2 -q^2 y^4$, and this can be positive for a range of $y$ if $k =-1$ and $q$ is sufficiently small. This is inconsistent with supersymmetry as the existence of a Killing spinor implies the existence of a causal Killing vector, which for a diagonal metric requires $g_{tt} \leq 0$. Thus, this is an example of the integrability conditions being necessary but not sufficient. We expect that we actually only have supersymmetry in the case studied in \cite{Klemm:2013eca}, where
\begin{equation} \label{susycond}
    \mu^2 = 4(e^2 + g^2) k, \quad e^2 = \lambda g^2.
\end{equation}

Supersymmetry thus requires $k=1$ or $k=0$, and the supersymmetric solutions are naked singularities, as  $H(y) <0$; there are no horizons. This might seem surprising as \cite{Ferrero:2020twa} find a solution with a regular horizon with $e=0$. However, in our parametrization, $e=0$ implies $\lambda=0$ (unless we set $g=0$, which would also imply $\mu=0$, reducing us to vacuum AdS). Setting $\lambda=0$, we indeed have a double root of $H(y)$ at $y=0$, but this is a limiting case of the C metric which falls outside of this general analysis; it is the limit in which the C metric reduces to the charged soliton. Consistent with our discussion, \cite{Ferrero:2020twa} find the supersymmetric solutions have naked singularities when $e \neq 0$. 

\subsection{Supersymmetry and branes}

If a theory has a supersymmetry with a local spinor parameter $\epsilon$, the supersymmetry of the theory will be preserved when we consider the theory on a manifold with boundary if $\epsilon$ is an eigenspinor of $\Gamma^n = n_\mu \Gamma^\mu$, $\Gamma^n \epsilon = \pm \epsilon$, where $n_\mu$ is the normal to the boundary \cite{vanNieuwenhuizen:2005kg}. Thus, the boundary can preserve at most half of the supersymmetry. This condition implies that the vector bilinear $V^\mu = \bar \epsilon \Gamma^\mu \epsilon$ constructed from the general spinor parameter is tangential to the boundary. It also implies that the scalar bilinear $f = \bar \epsilon \epsilon$ vanishes at the boundary, as
\begin{equation}
    \bar \epsilon \epsilon = \pm \bar \epsilon \Gamma^n \epsilon = \pm i \epsilon^\dagger \Gamma^0 \Gamma^n \epsilon = \mp i (\Gamma^n \epsilon)^\dagger \Gamma^0 \epsilon = - i \epsilon^\dagger \Gamma^0 \epsilon = - \bar \epsilon \epsilon. 
\end{equation}

Thus, a simple necessary condition for a boundary  to preserve the supersymmetry of a bulk solution is that the bilinear calculated from the Killing spinor satisfies $f=0$ at the boundary. The bilinears for the supersymmetric C metric solutions satisfying \eqref{susycond} were computed in \cite{Klemm:2013eca}. They find 
\begin{equation}
    f = \frac{QP}{Q^2+P^2} [m - (p-q) (Q^2+P^2)] = - \frac{eg}{A(e^2+g^2)}[\frac{\mu}{2} + (x+y) (e^2+g^2) ]
\end{equation}
Thus for the supersymmetric solutions with $e \neq 0$, $f$ does not vanish on the $x=0$ or the $y=0$ surface, and the branes we are considering break the supersymmetry.

\section{Action calculation}
\label{action}

We compute the Euclidean action of the charged C-metric solution with branes at $x=0$ and $y=0$. The Euclidean C-metric is
\begin{equation}
 \dd s_E^2=\frac{1}{A^2(x-y)^2} \left[-H(y)\dd\tau^2-\frac{\dd y^2}{H(y)} +\frac{\dd x^2}{G(x)}+G(x)\dd\phi^2\right].
 \label{eq:metric}
\end{equation}
We choose a gauge potential which is regular at the Euclidean horizon and at the axis,
\begin{equation}
 \cA_E=\frac{2}{A\ell_\star}
 \left[-ie(y-y_-)\dd\tau+g(x-x_+)\dd\phi\right].
 \label{eq:potential}
\end{equation}
Here $e$ and $g$ are the electric and magnetic charge parameters of the Lorentzian solution. The electric potential becomes imaginary under the continuation $t=-i\tau$.

Smoothness at $x=x_+$ and $y=y_-$ fixes
\begin{equation}
 \Delta\phi=\frac{4\pi}{|G'(x_+)|},\qquad
 \Delta\tau=\frac{4\pi}{|H'(y_-)|}.
 \label{eq:periods}
\end{equation}
The Euclidean solution has $0\leq x\leq x_+$ and $y_-\leq y\leq0$. We glue two copies of this region across the branes and regulate the asymptotic region near $(x,y)=(0,0)$ by placing a cutoff surface at
\begin{equation}
 z_c=x-y=\eps.
\end{equation}
Thus
\begin{equation}
 \mathcal M_\eps=\{0\leq x\leq x_+,\quad y_-\leq y\leq0,\quad x-y\geq\eps\}.
\end{equation}

%

The calculation of the gravitational part of the action is the same as in the uncharged case; the only difference is in the form of the functions $G(x)$ and $H(y)$. Including boundary counterterms and joint terms where the branes meet the boundary, the renormalized gravitational action is \cite{Hennigar:2026mxv}
\begin{equation}
 I_{E,\grav}^{\ren}
 =\frac{\Delta\tau\Delta\phi}{8\pi G_4A^2}
 \left[\frac{1+\lambda}{(x_+-y_-)^2}
 -\frac1{x_+^2}-\frac{\lambda}{y_-^2}\right].
 \label{eq:grav-ren}
\end{equation}
The root equations give the useful identity
\begin{equation}
 \frac1{x_+^2}+\frac{\lambda}{y_-^2}
 =(x_+-y_-)\left[\mu+q^2(x_++y_-)\right].
\end{equation}

\subsubsection*{Maxwell action}
For two copies of the wedge, the Euclidean Dirichlet action is
\begin{equation}
 I_M^D=\frac1{2g_\star^2}\int_{\mathcal M_\eps}\dd^4x\sqrt g\,F_{\mu\nu}F^{\mu\nu}
 =\frac{\ell_\star^2}{32\pi G_4}
 \int_{\mathcal M_\eps}\dd^4x\sqrt g\,F_E^2.
 \label{eq:maxwell}
\end{equation}
\begin{figure}
    \centering
    \includegraphics[width=0.75\linewidth]{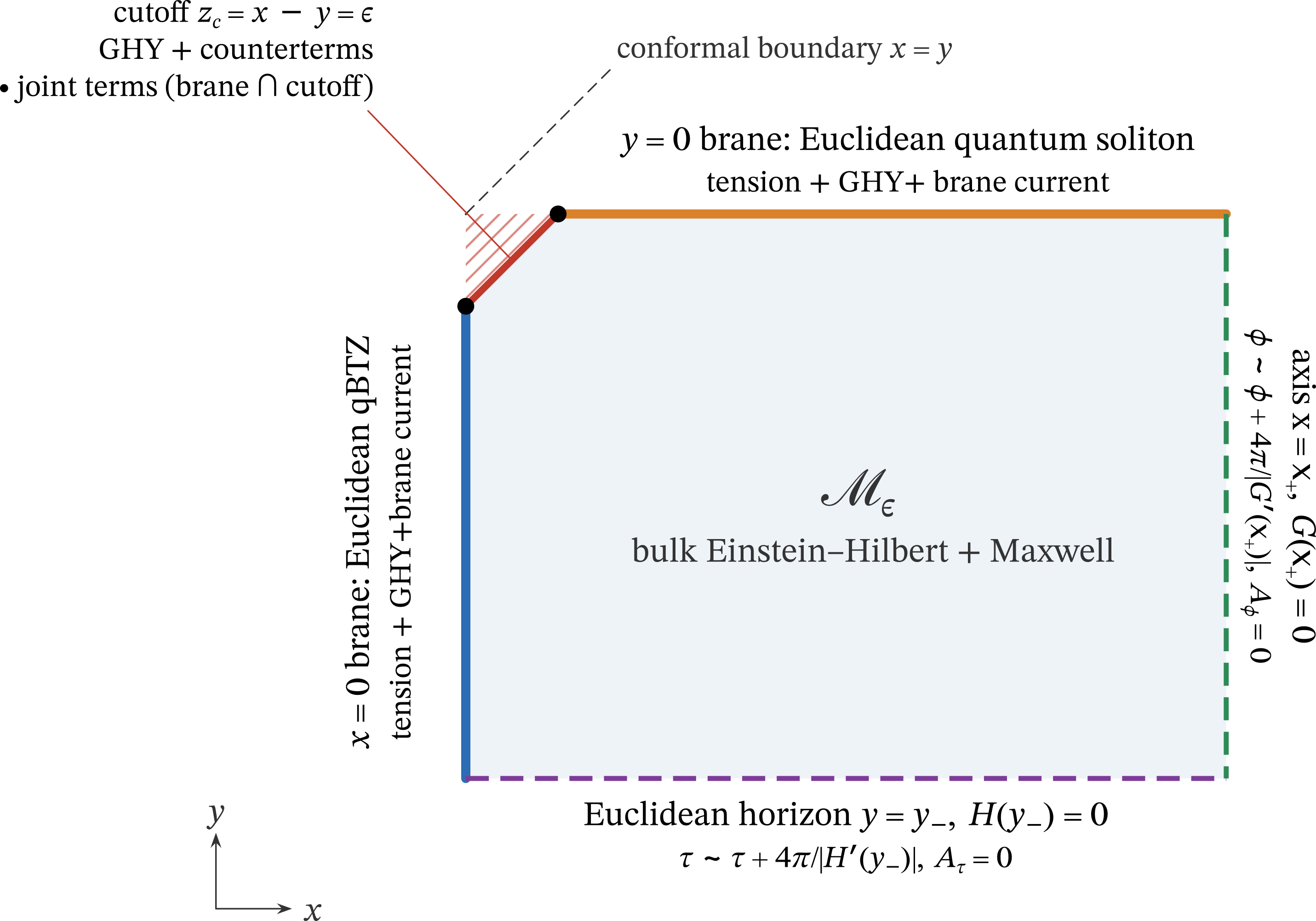}
    \caption{Schematic depiction of the two-brane action computation.}
    \label{fig:placeholder}
\end{figure}
On the charged solution, the field strength and its contraction are
\begin{align}
 F_E&=\frac2{A\ell_\star}(-ie\,\dd y\wedge\dd\tau+g\,\dd x\wedge\dd\phi),\notag\\
 F_E^2&=\frac{8A^2(x-y)^4}{\ell_\star^2}(g^2-e^2),&
 \sqrt g\,F_E^2&=\frac8{A^2\ell_\star^2}(g^2-e^2).
\end{align}
The powers of $x-y$ cancel in $\sqrt g F_E^2$, so the integrand is constant. We only need the coordinate area of the regulated region,
\begin{equation}
  \mathrm{area}_\eps
 =\int_0^\eps\dd x\int_{y_-}^{x-\eps}\dd y
 +\int_\eps^{x_+}\dd x\int_{y_-}^{0}\dd y
 =-x_+y_--\frac{\eps^2}{2}.
\end{equation}
Consequently,
\begin{equation}
 I_M^D(\eps)=\frac{\Delta\tau\Delta\phi}{8\pi G_4A^2}
 \left[2x_+y_-(e^2-g^2)+(e^2-g^2)\eps^2\right].
 \label{eq:maxwell-eval}
\end{equation}
The removed triangle contributes at order $\eps^2$. The Maxwell action is therefore finite as $\eps\to0$ and requires no counterterm.

We can also check whether the Maxwell action requires a joint term. Define the flux density on a boundary face $B$ by
\begin{equation}
 \pi_B^i=\sqrt{h_B}\,n_\mu F^{\mu i}.
 \label{eq:pi}
\end{equation}
The variation of \eqref{eq:maxwell} is
\begin{equation}
 \delta I_M^D=-\frac2{g_\star^2}
 \int_{\mathcal M_\eps}\dd^4x\sqrt g\,(\nabla_\mu F^{\mu\nu})\delta \mathcal{A}_\nu
 +\frac2{g_\star^2}\sum_B\int_B\dd^3x\,\pi_B^i\delta \mathcal{A}_i.
 \label{eq:maxwell-var}
\end{equation}
The boundary contribution vanishes when $\mathcal{A}_i$ is fixed on each face. Since it contains no derivative of $\delta \mathcal{A}_i$, there is no further integration by parts and no contribution at the joints. Varying the metric gives the same conclusion. For one copy of the bulk,
\begin{equation}
 \delta_g I_{M,s}^D=-\frac12\int_{\mathcal M_\eps^{(s)}}\dd^4x\sqrt g\,
 T_M^{\mu\nu}\delta g_{\mu\nu},\qquad
 T_M^{\mu\nu}=\frac1{g_\star^2}
 \left(F^{\mu\rho}F^\nu{}_\rho-\frac14g^{\mu\nu}F_{\rho\sigma}F^{\rho\sigma}\right),
\end{equation}
which contains no derivative of $\delta g_{\mu\nu}$ and produces no boundary term.


\subsubsection*{Brane currents}
The metric and gauge field are invariant under reflection across each brane. The outward Maxwell fluxes from the two bulk copies therefore add, and the branes must carry currents. The Lorentzian junction condition and the coupling to the gauge field are
\begin{equation}
 j_B^i=\frac2{g_\star^2}n_\mu F^{\mu i},\qquad
 I_B^{(j)}=\int_B\dd^3x\sqrt{-h_B}\,\mathcal{A}_i j_B^i.
 \label{eq:current-lor}
\end{equation}
We take the conserved current densities to be fixed sources, independent of the metric. Their coupling then adds no stress tensor, so the brane geometry continues to satisfy the Israel condition with tension alone. The action below applies to prescribed currents; including dynamical matter would also require its action and stress tensor.

After Wick rotation, the source coupling is
\begin{equation}
 I_{j,B}=-\int_B\dd^3x\,\mathcal{A}_i\mathcal J_B^i,\qquad
 \mathcal J_B^i=\frac2{g_\star^2}\pi_B^i
 \quad\hbox{on the solution}.
 \label{eq:current-eucl}
\end{equation}
Varying $\mathcal{A}_i$ at fixed $\mathcal J_B^i$ gives the Maxwell junction condition. Once this condition is imposed, the source action becomes
\begin{equation}
 I_{j,B}=-\frac2{g_\star^2}\int_B\dd^3x\,\mathcal{A}_i\pi_B^i.
 \label{eq:source-eval}
\end{equation}

On the $y=0$ brane,
\begin{equation}
 \pi_y^\tau=-\frac{2ie}{A\ell_\star},\qquad
 \mathcal{A}_\tau|_{y=0}=\frac{2iey_-}{A\ell_\star},
\end{equation}
and hence
\begin{equation}
 I_{j,y}(\eps)=\frac{\Delta\tau\Delta\phi}{8\pi G_4A^2}
 \left[-4e^2y_-(x_+-\eps)\right].
 \label{eq:source-y}
\end{equation}
On the $x=0$ brane,
\begin{equation}
 \pi_x^\phi=-\frac{2g}{A\ell_\star},\qquad
 \mathcal{A}_\phi|_{x=0}=-\frac{2gx_+}{A\ell_\star},
\end{equation}
so that
\begin{equation}
 I_{j,x}(\eps)=\frac{\Delta\tau\Delta\phi}{8\pi G_4A^2}
 \left[4g^2x_+(y_-+\eps)\right].
 \label{eq:source-x}
\end{equation}
The azimuthal current has the same sign as $\pi_x^\phi$. In the coordinates of the $x=0$ brane, with $\ell=1/A$ and $r=-\ell/y$, it is
\begin{equation}
 j_{B,x}^i=-\frac{4g\ell}{g_\star^2\ell_\star r^3}\delta_\phi^i.
 \label{eq:current-sign}
\end{equation}
The sign follows from the outward normal and the Maxwell junction condition.

The terms proportional to $\eps$ in \eqref{eq:source-y} and \eqref{eq:source-x} vanish when the cutoff is removed. A gauge transformation of the source action gives an edge integral proportional to $r_i\pi_B^i$, where $r_i$ is normal to the edge within the brane. The current on the $y=0$ brane is along $\tau$, and the current on the $x=0$ brane is along $\varphi$. Both are tangent to the edges, so
\begin{equation}
 r_i\pi_y^i=0,\qquad r_i\pi_x^i=0.
\end{equation}
The source action is therefore gauge invariant without an additional edge term.

\subsubsection*{Total action}
Adding the gravitational, Maxwell and brane current contributions, and taking $\eps\to0$, gives
\begin{equation}
 I_{E}
 =\frac{\Delta\tau\Delta\phi}{8\pi G_4A^2}
 \left[\frac{1+\lambda}{(x_+-y_-)^2}
 -\frac1{x_+^2}-\frac{\lambda}{y_-^2}
 -2(e^2-g^2)x_+y_-\right].
 \label{eq:total-src}
\end{equation}
This action includes both copies of the bulk and one source coupling on each physical brane.


\bibliographystyle{JHEP-2}
\bibliography{bib.bib}

\end{document}